%% file: main-arxiv.tex
\pdfoutput=1
\documentclass{article}
\usepackage[utf8]{inputenc}
\usepackage{stackengine}
\usepackage{amsfonts,amsmath,amssymb,amsthm}
\usepackage[letterpaper, portrait, margin=1in]{geometry}
\usepackage{verbatim,float,url,graphicx,psfrag}
\usepackage{natbib}
\usepackage{graphicx}
\usepackage{mathtools} 
\usepackage{cases} 
\usepackage{mathabx}   
\usepackage{hyperref} 
\usepackage{bbm}      
\usepackage{cancel}   
\usepackage[normalem]{ulem}  
\usepackage{mathtools} 
\usepackage{dsfont}
\usepackage{color}
\usepackage[export]{adjustbox} 
\usepackage{tabularx} 
\usepackage{tabulary}
\usepackage{marginnote}
\usepackage{multirow}
\usepackage{sistyle}
\SIthousandsep{,}
\input{title}
\usepackage{authblk}
\title{\ourtitle}
\author[1]{Sangwon Hyun\thanks{shyun2@ucsc.edu}}
\author[2]{Tim Coleman\thanks{tscoleman226@gmail.com}}
\author[3]{Francois Ribalet\thanks{ribalet@uw.edu}}
\author[2]{Jacob Bien\thanks{jbien@usc.edu}}
\affil[1]{Department of Statistics, University of California, Santa Cruz}
\affil[2]{Department of Data Sciences and Operations, University of Southern
  California}
\affil[3]{School of Oceanography, University of Washington}
\input{macros}

\begin{document}
\maketitle

\begin{abstract}
\input{abstract}

\end{abstract}

\let\clearpage\relax
\include{main-content}
\include{supp-content}

\bibliographystyle{apalike}
\bibliography{flowtrend}
\end{document}

%% file: title.tex
\newcommand{\ourtitle}{Trend Filtered Mixture of Experts for Automated Gating of High-Frequency Flow Cytometry Data}

%% file: macros.tex
\usepackage{booktabs}
\usepackage{subfig}
\usepackage{rotating} 
\usepackage{dsfont}
\usepackage[long]{optidef}

\def\one{\mathds{1}}

\def\E{\mathbb{E}}

\def\diag{\mathrm{diag}}

\def\R{\mathbb{R}}

\def\gh
\def\cI{\mathcal{I}}

\def\cN{\mathcal{N}}

\def\one{\mathds{1}}

\usepackage{pbox}

 \def\one{\mathds{1}}
\newcommand{\argmin}{\mathop{\mathrm{argmin}}}
\newcommand{\argmax}{\mathop{\mathrm{argmax}}}


%% file: abstract.tex

Ocean microbes are critical to both ocean ecosystems and the global climate. Flow cytometry, which measures cell optical properties in fluid samples,
 is routinely used in oceanographic research. Despite decades of accumulated data, identifying key microbial populations (a process known as ``gating'') remains a significant analytical challenge.
To address this, we focus on gating multidimensional, high-frequency flow cytometry data collected {\it continuously} on board oceanographic research vessels, capturing time- and space-wise variations in the dynamic ocean.   
Our paper proposes a novel mixture-of-experts model in which both the gating function and the experts are given by trend filtering.  
The model leverages two key assumptions: (1) Each snapshot of flow cytometry data is a mixture of multivariate Gaussians and (2) the parameters of these Gaussians vary smoothly over time.  
Our method uses regularization and a constraint to ensure smoothness and that cluster means match biologically distinct microbe types. We demonstrate, using flow cytometry data from the North Pacific Ocean, that our proposed model accurately matches human-annotated gating and corrects significant errors.

%% file: main-content.tex
\section{Introduction}

Flow cytometry is a technology used to quantify the abundance of different cell populations in a fluid sample. 
Cells pass single file through a laser, and a vector of optical characteristics of each cell is measured, which is used to infer the cell's subtype. 
While flow cytometry was originally developed for medical research, this technology was introduced to oceanography 40 years ago, leveraging the natural fluorescence of chlorophyll and other pigments within photosynthetic microbes to build valuable time-series data \citep{Sosik2010}.
Oceanographic flow-cytometers, like CytoBuoy \citep{Dubelaar1999-af}, FlowCytoBot \citep{Olson2003}, and SeaFlow \citep{seaflow-paper} can sample continuously, providing high-frequency data to study subtle, short-term fluctuations in microbial populations that are often missed by traditional, lower-frequency sampling methods. 
For instance, SeaFlow instruments have been deployed on over 100 research cruises in the last ten years, collecting data over 240,000 km in the Pacific Ocean, providing a broader and more detailed picture of microbial communities \citep{ribalet2019seaflow}.

Extracting meaningful information from flow cytometry data requires
identifying and isolating specific phytoplankton populations---a process known as ``gating"---from flow-cytometry data.
Traditionally, gating has relied on manual methods, with experts drawing boundaries around
populations based on visual interpretations of scatter plots and prior knowledge. However, manual gating is limited by subjectivity, inefficiency, environmental variability, scalability issues, and the inability to leverage high-dimensional data across large datasets.
To address the limitations of manual gating, machine learning approaches have become increasingly prevalent \citep{Cheung2021}. Gaussian mixture models (GMMs) have been used to probabilistically classify cells, assuming they originate from a mixture of Gaussian distributions. For time-series data, CYBERTRACK \citep{cybertrack} extends GMMs to model longitudinal cell population transitions and detect changes in mixture proportions .
However, these methods often assume constant optical properties, which is problematic for phytoplankton,
whose properties vary with environmental conditions.  
Also, an existing mixture-of-experts model called \texttt{flowmix} \citep{hyun2020modeling}, models the flow cytometry as a regression response to environmental factors. However, the most useful environmental factors 
are not always available in the ocean. Furthermore, their focus is in uncovering the regression relationship between the flow cytometry and changes in the environmental factors, not gating.

The focus of this paper is on the gating problem for a time series of flow cytometry data. Specifically, we would like to devise a method to partition cells into subtypes based on their optical
characteristics, for data in which the subtypes' mean optical characteristic and probabilities vary over time.  Although many pre-existing methods (\citealt{flowmeans, flowpeaks,
  flowsom}; more broadly reviewed by \citealt{Cheung2021}) have been designed for
traditional flow cytometry data (taken on a single fluid sample as in medical
applications), these methods do not translate well to the setting of continuous
flow cytometry data observed over time.  We develop a method that explicitly leverages the special
time-ordered structure of continuous flow cytometry data, without requiring external data such as environmental covariates.  A typical research
expedition takes days to weeks and can result in hundreds of millions of cells
being measured.  The standard gating pipeline involves a human data curator
making hundreds to thousands of cell-type determinations by eye.  Our method
allows for the automation of that process.  Doing so provides several advantages
over the manual gating approach, including increased efficiency, greater
consistency and reproducibility, and the ability to provide probabilistic
assignments of cells to subtypes in the case of overlapping subpopulations.

\begin{figure}[ht!]
\centering
\centering
  \includegraphics[width=.48\linewidth]{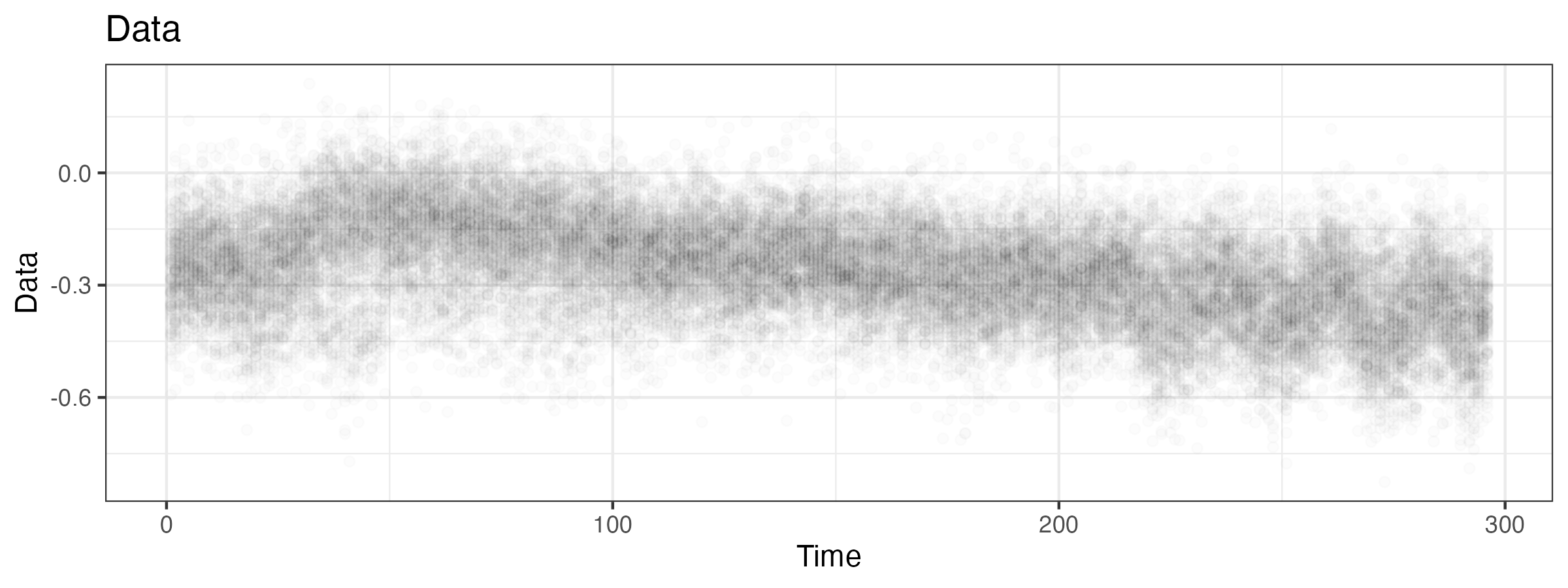}
  \includegraphics[width=.48\linewidth]{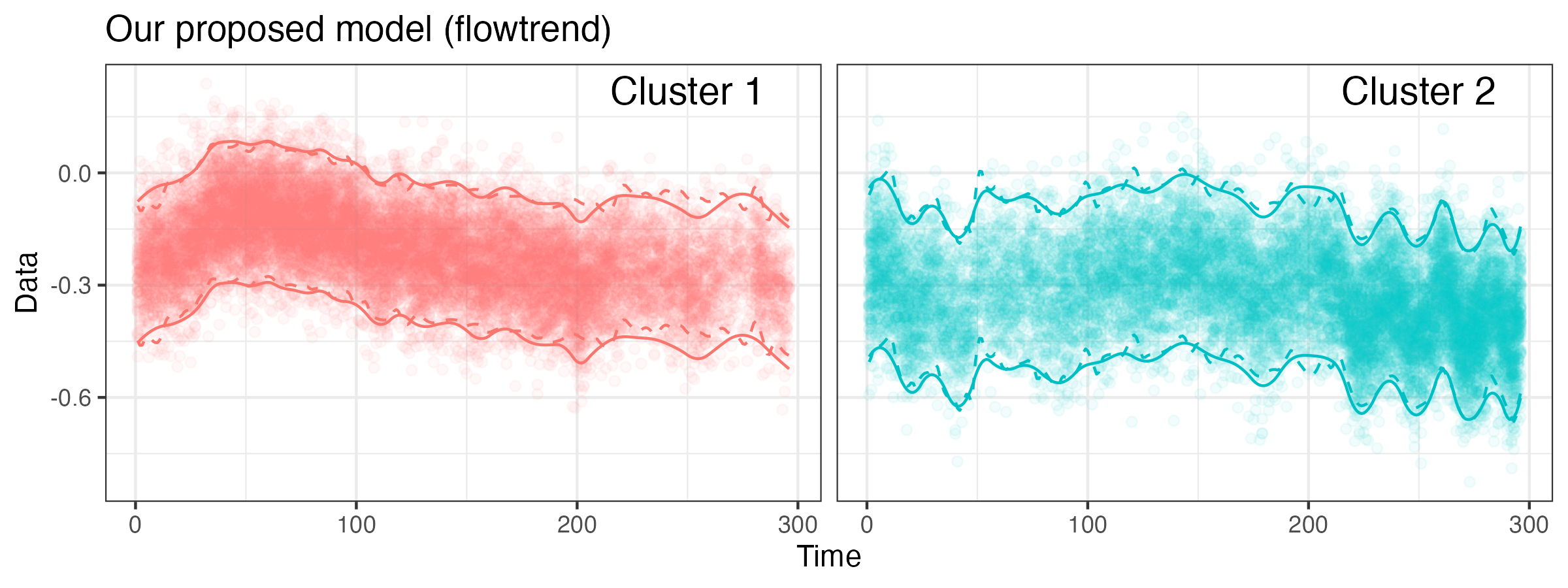}
  \includegraphics[width=.48\linewidth]{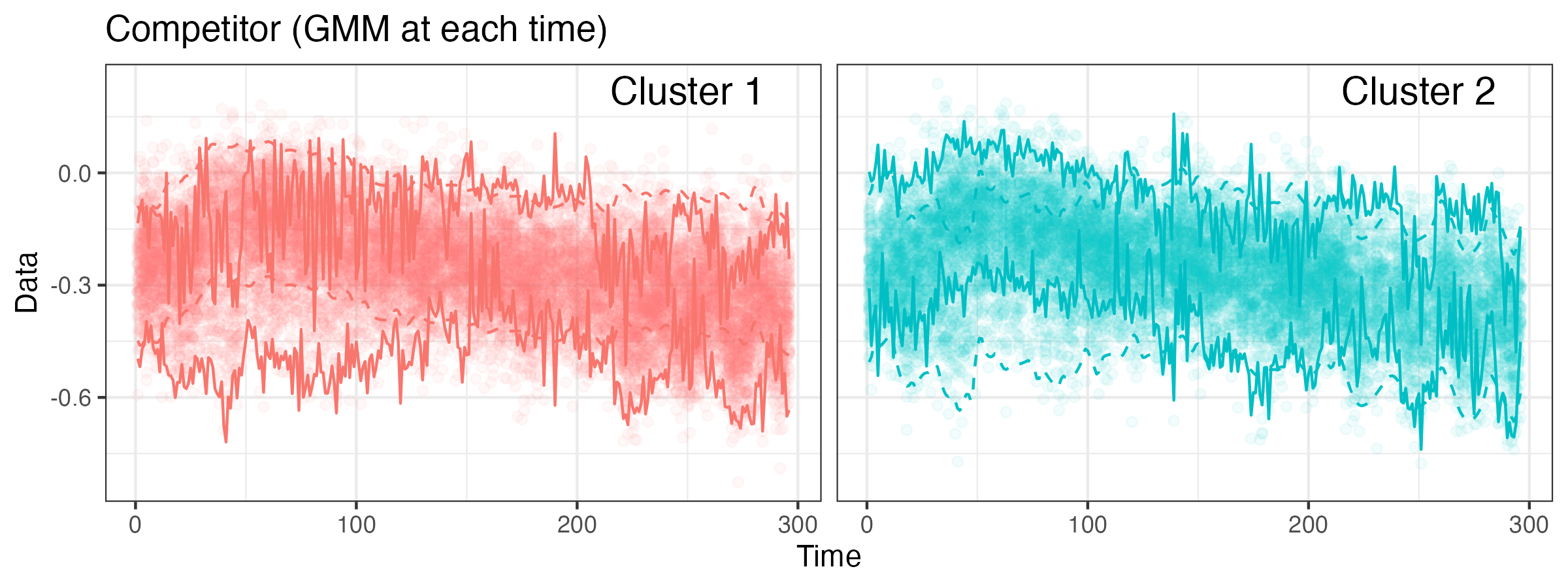}
  \includegraphics[width=.48\linewidth]{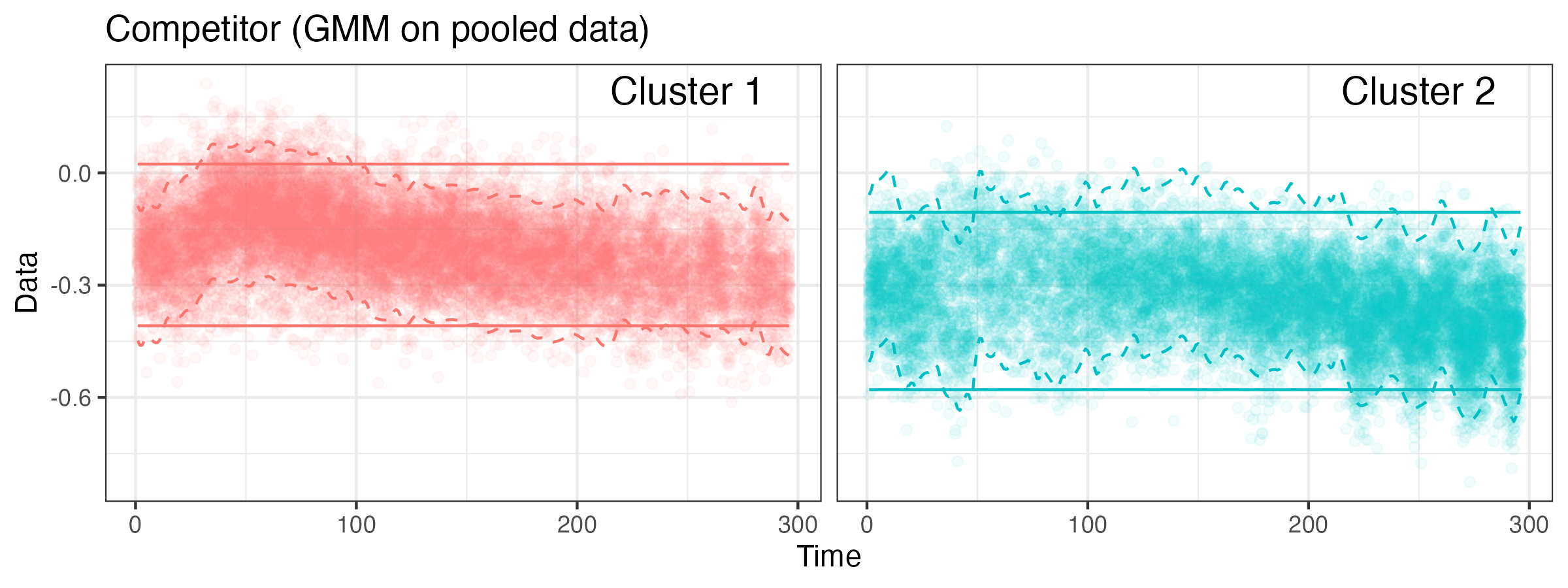}
  \caption{\it The top-left shows a pseudo-synthetic data example taken from a real dataset,
    for illustration.  Every vertical slice is a time point $t=1,\cdots, T$, and
    the $y$ axis is the value of a particle in a 1-dimensional flow cytometry
    dataset of size $n_t=100$ at time $t$. There are two Gaussian clusters at
    each time point, and the parameters -- means and probabilities -- of each
    cluster are changing gradually over time.  This is surprisingly hard to
    cluster accurately, and no tools are available for this problem. The first
    panel shows the data as-is. The remaining three panels show the result of
    ``gating'', or clustering the particles into two groups. The dashed lines
    show the true 95\% probability region for each cluster, and the solid lines
    show each model's estimated 95\% probability region from each cluster.  The
    {\tt flowtrend} model closely mimics the real model. The other two GMM-based
    baseline models are explained in Section~\ref{subsec:baseline_descrip}, and
    the full simulation is shown in
    Section~\ref{sec:NumericExperiments}. }
\label{fig:intro}
\end{figure}

To illustrate the challenge of the gating problem, the top-left panel of Figure~\ref{fig:intro} shows pseudo-synthetic
data. At each time point,
there are 100 data points (particles) generated from two-component mixtures of
Gaussians whose parameters and relative probabilities are derived from two
estimated gated cell populations---\textit{Prochlorococcus} and {\it Picoeukaryotes}---from
real flow cytometry data.  In this synthetic example, two populations are heavily overlapping so that
it is impossible to ascertain the two populations' parameters by eye.  It would be natural for a data analyst having access to a standard flow cytometry gating algorithm to consider one of two alternatives: (a) apply the algorithm to an aggregate of all cell data across all time (i.e., ignoring time information) or (b) apply the algorithm to each individual time, and then apply a matching algorithm to ensure a consistent meaning to cluster labels across time.  The bottom two panels of Figure~\ref{fig:intro} show the result of applying these two natural approaches (using a Gaussian
mixture model, GMM, as the base gating procedure).  We can see that both approaches are inadequate, since
their cluster estimates contain either too much variance or bias to be useful or
interpretable.

The top-right panel of Figure~\ref{fig:intro} shows the result of applying the method that we propose in this paper, which we call {\tt flowtrend}.  We can see that it is able to estimate the true underlying model quite accurately.  The key assumption made by {\tt flowtrend} is that the data
generating mechanism changes only gradually over time. Our method explicitly models
the smoothly changing nature of a time series of mixtures by combining a
nonparametric estimation technique with Gaussian mixture-of-experts modeling. In order to
induce smoothness, we employ a trend-filtering model \citep{steidl2006splines,
  kim2009ell_1, tibshirani2014adaptive} on the cluster means and probabilities
across time. And from the estimated Gaussian mixture models, we can assign
membership probabilities to each particle, which we can use for gating the
particles.

The most similar method is the \texttt{flowmix} mixture-of-experts model \citep{hyun2020modeling}, which connects the time points directly
by assuming the means and probabilities are regression functions of
environmental covariates such as salinity, temperature or nutrients. Because
many covariates are smooth across time, the means and probabilities are also
estimated to be gradually varying across time. The major downside of \texttt{flowmix} is that this
approach requires having environmental covariates at the analyst's disposal,
which is often not the case. Our proposed \texttt{flowtrend} model will be useful in the common setting that a scientist needs to perform gating but  does not have access to relevant environmental covariates or does not want to assume a functional
relationship between environment and cytograms.

Another method called CYBERTRACK \citep{cybertrack} uses Gaussian mixtures
to perform clustering for time-series flow cytometry data. However, CYBERTRACK only models the cluster probabilities $\pi$ over time as a stochastic process.  Crucially, CYBERTRACK does not allow for means to be different across time. Our proposed {\tt flowtrend} model is more appropriate for marine flow cytometry, where we know the cell populations' optical properties shift and oscillate over time. 
Also, our model restricts the mean movement over time, which serves as implicit regularization of the cluster locations in their overall variation, thus allowing for a more direct biological interpretation of the estimated subpopulations.

There is a literature on mixture of regressions that is tangentially
related. \cite{mixquantreg} studies a semi-parametric mixture of quantile
regressions model. \cite{Huang2013} considers a nonparametric mixture of
regressions, but it is only designed for a univariate response and covariate pair
and, furthermore, it is not designed for repeated responses.  \cite{Xiang2019} provides a
comprehensive overview of semi-parametric extensions of finite mixture models.
\cite{xiangyao2017} proposes a related method using single-index models. All of
these methods assume that covariates accompany a univariate, non-repeated
mixture responses. Our proposed method and software is explicitly designed for
multivariate and repeated responses that are gradually changing across time,
without the need for accompanying covariates.

Accompanying this manuscript, we also publish an \texttt{R} package called {\tt
  flowtrend} that implements our method and is designed to work on a
multi-thread high-performance parallel computing system such as SLURM
\cite{slurm}. We used literate programming \citep{Knuth1992-zz,litr} to write
{\tt flowtrend}, and the supplementary materials includes a bookdown
\citep{bookdown} which presents detailed explanations alongside the source code.

\section{Methodology}

In this section, we present our new method for modeling continuous flow cytometry data.  The first three subsections introduce the three main components of the optimization problem we will be solving: the data generating model underlying our method (Section~\ref{subsec:data_model}), the smoothness penalties which serve to regularize our estimates of the model parameters (Section~\ref{subsec:penalties}), and the parameter constraints (Section~\ref{subsec:restrict-mean}).  Section~\ref{subsec:Computation} presents an expectation-maximization algorithm based on the optimization problem.  Finally Sections~\ref{subsec:CV}~and~\ref{subsec:soft_gating} describe how to select tuning parameters and use the model outputs in gating.

\subsection{Basic model of cytogram data}
\label{subsec:data_model}

The data we will model consist of an ordered sequence of $d$-dimensional scatterplots (which are referred to as ``cytograms'' hereafter). The cytogram at time $t$ consists of
data points $y_{1}^{(t)},\ldots,y_{n_t}^{(t)} \in \R^d$, corresponding to the $n_t$ particles observed at time $t$.  Across the $T$ time points, there are a total of $N = \sum_{t=1}^T n_t$ points. (For applications without repeated responses, we can simply set
$n_t$ to $1$.) Our model assumes that each particle $y_i^{(t)}$ has a
latent cluster membership $Z_i^{(t)} \in \{1,\cdots, K\}$, which will be thought of as the particle's subtype.  The $K$ subtype relative abundances vary with time (both due to the ship's movement and due to the passage of time): $P(Z_i^{(t)} = k) = \pi_{kt}$.  Given a latent membership $Z_i^{(t)}$, we model $y_i^{(t)}$ to be
distributed as
\[
Y_i^{(t)} | \{Z_i^{(t)} = k\} \sim \mathcal{N}_d\left(\mu_{kt\cdot}, \Sigma_k \right).
\]
Here, the vector $\mu_{kt\cdot}\in\mathbb R^d$ represents cell subtype $k$'s
mean at time $t$, subsetted from $\mu \in\mathbb R^{K \times T \times
  d}$. Likewise, $\Sigma_{k} \in \mathbb R^{d \times d}$ is the $k$'th subtype's
variance-covariance. The choice to let the mean vary with time is based on the
time-varying nature of the $d$ optical properties being captured by flow
cytometry.  For example, the diameter and chlorophyll concentration of cells are
known to fluctuate on a daily cycle. Also, it is reasonable to assume the shape
of the distribution of a subtype, encoded in $\Sigma_k$ does not change over
time.

This mixture of Gaussians model has the following log-likelihood:
 \begin{equation}
    \label{eqn:ll_orig}
\ell\left(\{y_i^{(t)}\}_{i,t}; \mu, \Sigma, \pi\right) =  -\frac{1}{N}\sum_{t=1}^T\sum_{i=1}^{n_t} \log\left[\sum_{k=1}^K \pi_{kt} \cdot \phi\left(y_i^{(t)}; \mu_{kt\cdot}, \Sigma_k \right) \right],
\end{equation}
where $\Sigma$ and $\pi$ written without subscripts denote the collections of
parameters $\{\Sigma_{k}:1\le k \le K\}$, and
$\{\pi_{kt}:1\le k\le K,1\le t\le T\}$, respectively.

The particles in ocean flow cytometry data often have accompanying biomass estimates. Small particles with small biomass are likely to be more numerous than large particles. Let us call $C_i^{(t)}$ the biomass of particle $y_i^{(t)}$. To account for the relative importance that a large particle should have in the data, we reweight the log-likelihood of $y_i^{(t)}$ by a factor of $C_i^{(t)}$ so that a modified pseudo-likelihood is:
\begin{equation}
 \label{eqn:biomass_likelihood}
 \ell_\text{biomass}\left(\{{y_i^{(t)}}\}_{i,t} , \{C_i^{(t)}\}_{i,t} ; \mu, \Sigma,\pi \right) = 
 -\frac{1}{N}\sum_{t=1}^T\sum_{i=1}^{n_t} C_i^{(t)}\log\left[\sum_{k=1}^K  \pi_{kt} \cdot \phi\left({y_i^{(t)}}; \mu_{kt\cdot}, \Sigma_k \right) \right],
 \end{equation}
 where $N = \sum_{t=1}^T C_i^{(t)}$. Also, since cytograms often contain many points (with $n_t$ as large as $100,000$)
computation can be challenging and compressing the
data size by {\it binning} the particle-level cytogram data is convenient. 
We do this by discretizing the cytogram space into $B = D^d$ hyper-rectangles $\{E_b\}_{b=1}^B$,
where $D$ is the number of discrete values in the grid along each axis, and recording the number of particles in each grid cell at each time point $t$
according to
\[
C_b^{(t)} = \sum_{i = 1}^{n_t} C_i^{(t)} \one(y_i^{(t)} \in E_b), \ \text{for } b = 1, ..., B ; t = 1 ,..., T.
\]
Let $\{y_b\}_{b=1}^B$ be the bin centers, and define $N = \sum_{b=1}^B \sum_{t=1}^T C_b^{(t)}$. From here, we can write the likelihood of the binned data,
 \begin{equation}
  \label{eqn:binned_likelihood}
 \ell_\text{binned}\left(\{{y_b}\}_{b=1}^B , \{C_b^{(t)}\}_{b=1}^B ; \mu, \Sigma,\pi \right) = 
 -\frac{1}{N}\sum_{t=1}^T\sum_{b=1}^{B} C_b^{(t)}\log\left[\sum_{k=1}^K  \pi_{kt} \cdot \phi\left(y_b; \mu_{kt\cdot}, \Sigma_k \right) \right] .
 \end{equation}
 To be clear, the
 pseudo-likelihoods (\ref{eqn:biomass_likelihood}) and \eqref{eqn:binned_likelihood} are
a step removed from the data generating process
 of the data; however, this approach is useful in practice for lessening class imbalance and making computation more feasible, and it has been used before \citep{hyun2020modeling}. The
 computational advantage of \eqref{eqn:binned_likelihood} is sizeable since the computation of our model depends on the number of terms in the sum of the pseudolikelihood, and in general $N\gg T\cdot B$, which in turn can be much smaller than the number of nonempty bins, $\sum_{t=1}^T \sum_{b=1}^B\one\{C_b > 0\}$. Bin sparsity leads to
 considerable memory savings as well: for example, for a dataset
 with $T=296$ (considered in Section~\ref{sec:3dreal}) the original particle-level data are several gigabytes whereas the binned data are in the tens of megabytes.  Lastly, the expression in
 \eqref{eqn:biomass_likelihood} is easily reduced to the true model when
 particles $y_{i}^{(t)}$ are observed without observation weights like biomass
 simply by taking $C_i^{(t)} = 1$. We use the pseudolikelihood in
 \eqref{eqn:biomass_likelihood} as the basis of the objective value to optimize
 for the rest of the paper because it subsumes \eqref{eqn:binned_likelihood} as a special case.

\subsection{Trend filtering of each cluster's \texorpdfstring{
$\mu$}{mean} and \texorpdfstring{$\pi$}{probabilities}}
\label{subsec:penalties}

The basic model in Section~\ref{subsec:data_model} places no restrictions on $\mu_{kt\cdot}$ and
$\pi_{kt}$, which allows them to be highly varying and
choppy over time. On the other hand, it is apparent from plotting ocean flow
cytometry data that the cell populations are slowly varying in time.  In order to incorporate this knowledge while estimating
model parameters $\mu_{kt\cdot}$ and $\pi_{kt}$, we make use of \textit{trend filtering}, reviewed next.

Trend filtering \citep{steidl2006splines, kim2009ell_1} is a tool for
non-parametric regression on a sequence of output points $v = (v_1,..., v_T)$
observed at locations $x = (x_1, ..., x_T)$.  If $x$ is equally spaced, then the $l$-th order trend filtering
estimate is obtained by solving
\begin{equation*}
\min_{\psi \in \mathbb{R}^T} \frac{1}{2}\| v-\psi\|_2^2 +
  \lambda \| D^{(l_\mu+1)} \psi\|_1,
\end{equation*}
where $\lambda$ is a tuning parameter and
$D^{(l_\mu+1)} \in \R^{(T-l-1) \times l}$ is the $(l_\mu+1)$-th order discrete differencing
matrix defined recursively as $D^{(l+1)} = D^{(1)} D^{(l)}$, starting with $D^{(1)} \in \R^{(T-1) \times T}$:
\begin{equation*}
    D^{(1)} = 
    \begin{bmatrix}
    -1 & 1 & 0 & \cdots & 0 & 0 \\
    0 & -1 & 1 & \cdots & 0  & 0\\
    \vdots & & & & \\
    0 & 0 & 0 & \cdots & -1 & 1
    \end{bmatrix}.
\end{equation*}
An order-$l$ trend filtering mean estimate is a
piecewise polynomial of order $l$. For example, $l = 0$ will produce a piecewise
constant mean estimate $\hat \mu$.  This special case is better known as the ``fused lasso'' (with no sparsity penalty, \citealt{tibshirani2005sparsity})  Likewise, $l = 1$ will produce
a piecewise linear estimate, while $l=2$ will produce a piecewise quadratic estimate.  A straightforward change in $D^{(l)}$ allows for the input $x$ to
be unevenly spaced (\citealt{tibshirani2014adaptive}, Section 6).

Porting these ideas to our problem, we add trend filtering penalties to encourage smoothness (with respect to time) in each $\mu_{k\cdot\cdot}\in\mathbb R^{T\times d}$ and $\pi_{k\cdot}\in\mathbb R^{T}$.  For $\mu_{k\cdot\cdot}$, we use
the penalty $\lambda_\mu \|D^{(l_\mu+1)}\mu_{k\cdot\cdot}\|_1$. Here, we use the
notation that for a matrix $M$, $\|M\|_1=\|\text{vec}(M)\|_1$.  For
$\pi_{k\cdot}$, we apply the penalty on the scale of logits $\alpha_{kt}$
(denoting the $(k,t)$'th entry of $\alpha \in \R^{K \times T}$), where
\begin{equation}
  \label{eqn:pi-alpha}
  \pi_{kt} = \frac{\exp(\alpha_{kt})}{\sum_{m=1}^K \exp(\alpha_{mt})}.
\end{equation}
Putting this
together, we aim to minimize the following penalized negative log-likelihood:
\begin{equation}
  \label{eqn:pen_likelihood}
- \ell_\text{biomass}\left(\{{y_i^{(t)}}\}_{i,t} , \{C_i^{(t)}\}_{i,t} ; \mu, \Sigma,\pi \right)
  + \lambda_\mu \sum_{k=1}^K \| D^{(l_\mu + 1)} \mu_{k\cdot\cdot}\|_1 + \lambda_\pi \sum_{k=1}^K \| D^{(l_\pi +1)} \alpha_{k\cdot}\|_1,
\end{equation}
where $\lambda_\mu, \lambda_\pi$ are tuning parameters and $l_\mu, l_\pi$ are
the degree of trend filtering for $\{\mu_{kt\cdot}\}_{k,t}$ and $\{\pi_{kt}\}_{k,t}$ respectively.

\subsection{Restricting mean movement}
\label{subsec:restrict-mean}

It is important in our application that each mean trajectory correspond consistently across all times to the same biological cell population.
While the trend filtering penalties in the previous section help encourage this behavior, 
we wish to enforce this more directly.
Cells from the same microbial type cannot change too much over time in their
characteristics, being bound by the bio-physiological limits of a single subspecies.  

We therefore 
add to \eqref{eqn:pen_likelihood} an explicit constraint, requiring that all the mean vectors for cluster $k$ across time, $\{\mu_{kt\cdot}:t=1,\ldots,T\}$, remain within a radius $r$ of their time average $\bar\mu^{(k)} = 1/T\cdot\sum_{t=1}^T \mu_{kt\cdot}$: 

 \begin{mini}{
    \substack{\mu, \Sigma, \alpha }}{ -\frac{1}{N}\sum_{t=1}^T\sum_{i=1}^{n_t} C_i^{(t)}\log\left[\sum_{k=1}^K \pi_{kt} \cdot \phi\left(y_i^{(t)}; \mu_{kt\cdot}, \Sigma_k \right) \right] + \lambda_\mu \sum_{k=1}^K \| D^{(l_\mu + 1)} \mu_{k\cdot\cdot}\|_1 + \lambda_\pi \sum_{k=1}^K \| D^{(l_\pi +1)} \alpha_{k\cdot}\|_1}{}{}
  \addConstraint{\| \mu_{ kt\cdot} - \bar{\mu}^{(k)}\|_2 \le r \;\;\forall t=1,\cdots, T, \text{and} \  \forall k = 1, \cdots, K}. 
\label{eqn:pen_likelihood_ball}
\end{mini}
A similar constraint was imposed in \citet{hyun2020modeling} for the same 
reason. In contrast to $\lambda_\pi$ and $\lambda_\mu$
which are tuning parameters controlling the amount of smoothness, the value of $r$ can be reasonably
pre-specified using subject matter knowledge about the variability of
optical properties of cell populations of interest.
 
\subsection{Expectation-maximization (EM) algorithm}
\label{subsec:Computation}
The mixture likelihood in  \eqref{eqn:pen_likelihood_ball} makes this a non-convex
function of $\mu, \Sigma$, and $\pi$. We therefore employ an EM \citep{Dempster1977} approach, which iteratively breaks this into simpler problems to solve.  Following the standard strategy, we imagine augmenting the data with latent cluster memberships,
$Z_i^{(t)}$, for each particle $y_i^{(t)}$, and write the log-likelihood $\ell_c$ of the
\textit{complete} data, corresponding to the joint distribution of
$\{Z_i^{(t)}\}_{i,t}$ and $\{y_i^{(t)}\}_{i,t}$:
\[
\ell_c\left(\{y_i^{(t)}\}_{i,t}, \{Z_i^{(t)}\}_{i,t}; \mu, \Sigma, \pi\right) =  -\frac{1}{N}\sum_{t=1}^T\sum_{i=1}^{n_t} \sum_{k=1}^K \one(Z_i^{(t)} = k) \cdot \left[\log\pi_{kt} + \log\phi\left(y_i^{(t)}; \mu_{kt\cdot}, \Sigma_k \right) \right].
\]
Because the latent cluster memberships $\{Z_i^{(t)}\}_{i,t}$ are unobservable, we integrate them out to form a surrogate objective function. For a set of
parameters $\tilde{\theta} = (\tilde{\pi}, \tilde{\mu}, \tilde{\Sigma})$, we take the conditional expectation of the log-likelihood, given the observed data, $\E_{\tilde{\theta}}[\ell_c| \{y_i^{(t)}\}]$, over the latent variables:
\begin{equation}
\label{eqn:surrogate_objective0}
-\frac{1}{N}\sum_{t=1}^T\sum_{i=1}^{n_t} \sum_{k=1}^K \tilde{\gamma}_{itk} \cdot \left[\log\pi_{kt} + \log\phi\left(y_i^{(t)}; \mu_{ k t\cdot} , \Sigma_k \right) \right],
\end{equation}
where
$\tilde{\gamma}_{itk} = \E_{\tilde{\theta}} [\one(Z_i^{(t)} = k) \mid y_i^{(t)}]
= \mathbb P_{\tilde\theta}(Z_i^{(t)} = k \mid y_i^{(t)})$ is the conditional membership probability
of particle $y_i^{(t)}$ to cluster $k$ assuming a set of parameters
$\tilde{\theta}$. These are sometimes called responsibilities in the literature, a
terminology we will also adopt. We use \eqref{eqn:surrogate_objective0} as a
surrogate objective since it is easier to optimize, being convex in $\pi$ and biconvex in $\mu$ and each $ \Sigma_k^{-1})$. With weighted data, the surrogate (unpenalized) objective for a given parameterization
$\tilde{\theta}$ is simply
\begin{equation}
\label{eqn:surrogate_binned}
-\frac{1}{N}\sum_{t=1}^T\sum_{i=1}^{n_t} C_i^{(t)} \sum_{k=1}^K \tilde{\gamma}_{itk} \cdot \left[\log\pi_{kt} + \log\phi\left({y}^{(t)}_i; \mu_{kt\cdot} , \Sigma_k \right) \right].
\end{equation}
Given the latest responsibilities
$\tilde \gamma = \{\tilde \gamma_{itk}\}_{i,t,k}$, we will optimize the following
over $\mu$, $\Sigma$, and $\pi$ instead of solving \eqref{eqn:pen_likelihood_ball}:
\begin{multline}
Q_{\tilde{\theta}} (\mu, \Sigma, \pi) = -\frac{1}{N}\sum_{t=1}^T\sum_{i=1}^{n_t} C_i^{(t)} \sum_{k=1}^K \tilde{\gamma}_{itk} \cdot \left[\log\pi_{kt} + \log\phi\left(y_i^{(t)}; \mu_{k t \cdot} , \Sigma_k \right) \right] + \\ \lambda_\mu \sum_{k=1}^K \| D^{(l_\mu + 1)} \mu_{k \cdot \cdot}\|_1 + \lambda_\pi \sum_{k=1}^K \| D^{(l_\pi +1)} \alpha_{k\cdot}\|_1 + \sum_{t=1}^T\sum_{k=1}^K \one_\infty(\| \mu_{ k t \cdot} - \bar{\mu}^{(k)} \|_2 \leq r),
\label{eqn:surrogate_objective}
\end{multline}
where $\one_\infty\{\cdot\}$ is an infinite indicator that encodes hard constraints.
We optimize a sequence of such surrogates in a penalized EM algorithm---to be
described next---by alternating between the following two steps:
\begin{enumerate}
\item E-step: Estimate the particle
  responsibilities $\tilde{\gamma}$.
\item M-step: Estimate $(\mu,\Sigma,\pi)$ the parameters by maximizing
  the surrogate objective \eqref{eqn:surrogate_objective}.
\end{enumerate}
Prior to applying the algorithm, the initial values of the parameters are set as
follows. The inital cluster means $\mu_{kt\cdot}$ are chosen to be the same for all
$t$, i.e. $\mu_{kt\cdot} = m_k$, where $m_k \in \R^d$ is randomly sampled from the empirical
distribution of all $\{y_i^{(t)}\}_{i,t}$ ({\it after} truncating the heights so
that the high-density regions are capped). This strategy is designed to prevent
initial cluster means from only being in high-density regions in the data. The
cluster probabilities $\pi_{kt}$ are initialized as $1/K$ for all $k$ and $t$,
and the covariance matrices are initialized as identity matrices of dimension
$d \times d$.

Next we give details of the E- and M-steps.

\subsubsection*{E-Step}

For a given set of parameters $(\mu, \Sigma,\pi)$, we can calculate the
responsibility of a particle $y_i^{(t)}$ as a ratio of weighted Gaussian density
functions:
\begin{equation}
\label{eqn:resp_def}
\tilde \gamma_{itk}(\mu, \Sigma, \pi) = \frac{\pi_{kt} \cdot  \phi\left(y_i^{(t)}; \mu_{kt \cdot}, \Sigma_k\right)}{\sum_{j = 1}^K \pi_{jt} \cdot  \phi\left(y_i^{(t)}; \mu_{jt \cdot}, \Sigma_j\right)}.
\end{equation}

\subsubsection*{M-Step}

Once the responsibilities are estimated, we can derive updates for $(\mu, \Sigma, \pi)$ by optimizing
\eqref{eqn:surrogate_objective}  with respect to each parameter.
\begin{description}
\item[Updating $\mu$:] For fixed $\gamma$,
  we solve the following problem:
 \begin{multline}
   \label{eqn:min_mu_overall}
   \min_{\mu \in \mathbb{R}^{d \times T \times K}} -\frac{1}{N}\sum_{t=1}^T\sum_{i=1}^{n_t} C_i^{(t)} \sum_{k=1}^K \tilde{\gamma}_{itk} \cdot \left[\log\pi_{kt} + \log\phi\left(y_i^{(t)}; \mu_{k t\cdot } , \Sigma_k \right) \right] + \\ \lambda_\mu \sum_{k=1}^K \| D^{(l_\mu + 1)} \mu_{k \cdot \cdot}\|_1 + \sum_{t=1}^T\sum_{k=1}^K \one_\infty(\| \mu_{k t \cdot} - \bar{\mu}^{(k)} \|_2 \leq r).
 \end{multline}
 
 Importantly, the problem in \eqref{eqn:min_mu_overall} separates over
 $k$, so we can treat each cluster separately. Then, for cluster $k$, the problem becomes to
 solve,
 \begin{mini}{ \substack{\mu_{k \cdot \cdot} \in \mathbb{R}^{T \times d}}}{\frac{1}{2N}
     \sum_{t=1}^T \sum_{i=1}^{n_t} \tilde{\gamma}_{itk} (y_i^{(t)} - \mu_{kt \cdot})^\top \hat{\Sigma}_k^{-1} ( y_i^{(t)} - \mu_{kt \cdot}) + \lambda_\mu
     \sum_{j=1}^d \|D^{(l_\mu+1)}\mu_{k\cdot j}\|_1}{}{}
 \addConstraint{\| \mu_{kt \cdot} - \bar{\mu}^{(k)}\|_2 \le r \;\;\forall t=1,\cdots, T, }\label{eq:mstep-objective}
\end{mini}

To solve this minimization
problem, we devise an alternating direction method of multipliers (ADMM; \citealt{boyd-admm})
algorithm. We first reformulate \eqref{eq:mstep-objective} as follows:

\begin{mini}{
    \substack{\mu, w, z}}                         {\frac{1}{2N}
    \sum_{t=1}^T \sum_{i=1}^{n_t} \tilde{\gamma}_{itk} (y_i^{(t)} - \mu_{kt\cdot})^\top \hat{\Sigma}_k^{-1} ( y_i^{(t)} - \mu_{kt\cdot})
  + \lambda_\mu \sum_{j=1}^d \|D^{(1)}w_{\cdot j}\|_1}{}{}
  \addConstraint{\| z_{t \cdot} \|_2 \le r \;\;\forall t=1,\cdots, T} 
   \addConstraint{D^{(l_\mu)}\mu_{k \cdot j} = w_j \;\; \forall j = 1, \cdots, d} 
  \addConstraint{ \mu_{kt\cdot} - \bar{\mu}^{(k)} = z_{t} \;\; \forall t = 1, \cdots, T}. 
\label{eq:mstep-objective-augmented}
\end{mini}
Note that $z$ is in $\mathbb{R}^{T\times d}$, and $z_{t\cdot}\in\R^d$ is a column
vector taken from the $t$-th row of $z$. Also, $w \in \mathbb{R}^{(T-l) \times d}$ and
$w_{\cdot j}$ is a column vector taken from the $j$-th column of $w$. Then, introducing
the auxillary variables $\{u_z^{(t)}\in\R^d\}_{t=1}^{T}$ and
$\{u_w^{(j)} \in \R^{T - l}\}_{j=1}^d$, the augmented Lagrangian is given by:
\begin{align*}
  L(\mu_{k\cdot\cdot}, w,  z, \{u_w^{(j)}\}_1^d,
  \{u_z^{(t)}\}_1^T)
  =& 
     \frac{1}{2N} \sum_{t=1}^T
     \sum_{i=1}^{n_t} \tilde{\gamma}_{itk} (y_i^{(t)} - \mu_{k t \cdot})^\top \hat{\Sigma}_k^{-1} ( y_i^{(t)} - \mu_{k t \cdot}) \\
   &+ \lambda_\mu \sum_{j=1}^d \|D^{(1)}w_{\cdot j}\|_1 + \sum_{t=1}^T \one_\infty(\|z_{t\cdot}\|_2 \le r)\\
   &+ \sum_{t=1}^T \left[ {u_z^{(t)}}^\top ( \mu_{k t \cdot} - \bar{\mu}^{(k)} -z_{t \cdot} )
     + \frac{\rho}{2} \|\mu_{k t \cdot} - \bar{\mu}^{(k)} - z_{t \cdot}  \|^2 \right]\\
   & + \sum_{j=1}^d \left[ u_w^{(j)^\top}(D^{(l_\mu)}\mu_{k \cdot j} - w_{\cdot j} ) + \frac{\rho}{2} \|D^{(l_\mu)}\mu_{k \cdot j} - w_{\cdot j} \|^2 \right].
\end{align*}

From this, we can write down the ADMM updates: 

\begin{enumerate}
\item[1.] $\hat\mu_{k\cdot\cdot} \leftarrow \argmin_{\mu_{k \cdot \cdot}}
  L(\mu_{k\cdot\cdot}, \hat w,  \hat z, \{u_w^{(j)}\}_1^d, \{u_z^{(t)}\}_1^T)$
\item[2a.] $\hat z_{t \cdot} \leftarrow \argmin_{z_{t \cdot}} L(\hat\mu_{k\cdot\cdot}, \hat w,  z, \{u_w^{(j)}\}_1^d, \{u_z^{(t)}\}_1^T)$ for $t=1,\cdots, T$
 \item[2b.] $\hat w_j \leftarrow \argmin_{w_j} L(\hat\mu_{k\cdot\cdot}, w,  \hat z, \{u_w^{(j)}\}_1^d, \{u_z^{(t)}\}_1^T)$  for $j=1,\cdots, d$
 \item[3a.] $u_z^{(t)} \leftarrow u_z^{(t)} + \rho ( \hat\mu_{k t \cdot} - \bar{\mu}^{(k)} - \hat z_{t \cdot} )$ for
   $t=1,\cdots, T$ 
 \item[3b.] $u_w^{(j)} \leftarrow u_w^{(j)} + \rho (D^{(l_\mu)}\hat\mu_{k \cdot j} - \hat w_j)$ for $j= 1, \cdots, d$.
 \end{enumerate}
    
 Details for each of these ADMM updates are given in the
 Appendix~B.
 A notable characteristic of our algorithm is its ability to leverage (a) existing trend filtering solvers and (b) efficient Sylvester equation solvers.
 
\item[Updating $\pi$:] Smooth estimation of $\pi$ requires trend filtering of
  $\alpha_{k\cdot}\in\mathbb R^T$ for each cluster $k$, where we recall from
  \eqref{eqn:pi-alpha} that
  $\pi_{kt} = \frac{\exp(\alpha_{kt})}{\sum_{m=1}^K \exp(\alpha_{mt})}$.  While trend-filtering software  exists for the binomial and Poisson families (\texttt{glmgen},
  \citealt{arnold2014glmgen}), it does not exist for the multinomial family. We
  therefore re-express the minimization problem in \eqref{eqn:surrogate_objective} over
  $\alpha$ to a multinomial lasso problem, which we solve using $\texttt{glmnet}$ \citep{friedman2022package}. In particular, 
  replacing $\pi_{kt}$ in \eqref{eqn:surrogate_objective} with the corresponding expression involving $\alpha_{\cdot t}$,
we wish to solve
\begin{equation}
    \label{eqn:alpha_problem}
    \min_{\alpha} - \frac{1}{N} \sum_{t=1}^T \left\{\sum_{k=1}^K \tilde{\gamma}_{tk} \alpha_{kt}  - n_t \log \sum_{k=1}^K e^{\alpha_{kt}}\right\}  + \lambda_\pi \sum_{k=1}^K \| D^{(l_\pi +1)} \alpha_{k\cdot}\|_1,
\end{equation}
where   $\tilde \gamma_{tk} = \sum_{i=1}^{n_t}C_i^{(t)} \tilde \gamma_{itk}$. 
This is a multinomial regression with $\{\tilde \gamma_{tk}\}_{t,k}$ as responses and an
identity design matrix, together with a trend filtering penalty on coefficients
$\alpha_{k\cdot}$. While there is not a unique minimizer in terms of $\alpha$ (adding a constant vector achieves the same objective value),
the problem has a unique minimizer in terms of $\pi$). Using Lemma~2 from
\citet{tibshirani2014adaptive}, we transform \eqref{eqn:alpha_problem} to the
following ${\ell}_1$-penalized multinomial regression problem,
\begin{equation}
  \label{eqn:lasso_alpha_problem}
  \min_{\omega} -\frac{1}{N}\sum_{t=1}^{T}\left\{\sum_{k=1}^{K}\tilde{\gamma}_{tk}\omega^\top_kh_t- n_t\log\sum_{l=1}^Ke^{\omega_l^\top h_t}\right\}   + \frac{l_\pi !}{N^{l_\pi}} \lambda_\pi\sum_{k=1}^K \sum_{j=l_\pi + 2}^T | \omega_{kj}|
\end{equation}

where $\alpha_{kt} = h_t^\top\omega_{k}$, and $h_t$ are rows of a $T \times T$ trend
filtering regression matrix $H$ defined in
Appendix~C.
This optimization problem is an ${\ell}_1$-penalized multinomial regression, with no penalty on the first
$l_\pi + 1$ terms. Intuitively, $H$ consists of basis functions that are
polynomials of order $l_\pi$ that are 0 up until a specified point in time. We
use \texttt{glmnet} to regress responses $\{\tilde \gamma_{tk}\}_{t,k}$ against
regressors $\{h_t\}_t$ in a penalized multinomial regression.

\item[Updating $\Sigma$:]

 $\Sigma$ is updated by a weighted empirical covariance matrix,
\[
\Sigma_k \leftarrow \frac{\sum_{t=1}^T \sum_{i=1}^{n_t} C_{i}^{(t)} \tilde{\gamma}_{itk} (y_i^{(t)} - \mu_{kt\cdot} )(y_i^{(t)} -\mu_{kt\cdot})^\top}{\sum_{t=1}^T \sum_{i=1}^{n_t} \tilde{\gamma}_{itk}}. 
\]
\end{description}

\subsection{Prediction and cross validation}
\label{subsec:CV}

We next describe a cross validation approach for selecting the regularization parameters
$\lambda_\pi$ and $\lambda_\mu$. Since the data inputs to trend filtering are a
time series, we use a specialized split of the data for cross validation
following the $M$-fold cross validation described in the \texttt{cv.trendfilter}
function in the \texttt{genlasso} package \citep{arnold2020package}. Let $M$
denote the number of data folds. Set aside the endpoints $t=1$ and $t=T$; these are not
included in any fold but are later made available to every fold for prediction.
Next, every $M^{\text{th}}$ time point starting with time point $m$ is assigned
to fold $m$ (for $m=1,\ldots,M$), which we will call $I_m$. The complement is called
$I_{-m} := \bigcup_{(1:M)\backslash m} I_m$.  Also denote as
$y_{-m}:= \{y^{(t)}, t \not\in I_m\}$ the data not in fold $m$.
Then, a cross validation score is calculated via
\begin{equation}
  \label{eqn:CVll}
  CV_M(\lambda_\mu, \lambda_\pi) = \frac{1}{M} \sum_{m=1}^M \ell_{\text{biomass}}(\{y^{(t)}\}_{t \in I_m}
  , \{C^{(t)}\}_{t \in I_m}; \hat \mu^{(-m)}(I_m), \hat \pi^{(-m)}(I_m), \hat \Sigma^{(-m)}).
\end{equation}
Here, the out-of-sample mean estimates $\hat \mu^{(-m)}(I_m)$ are made (a) using
the estimated model $\hat \mu^{(-m)}(\cdot)$ from data $y_{-m}$, and (b)
linearly interpolated at the time points $I_m$.

Specifically, for cluster $k$, in order to compute an out-of-sample mean
estimate $\hat \mu^{(-m)}_{k} (\{t^*\})$ at a point $t^* \in I_m$, we first let
$\left\lfloor t^* \right\rfloor$ denote the nearest $t$ in $I_{-m}$ less
than $t^*$, and similarly define $\left\lceil t^* \right\rceil$. Then, we
compute:
\begin{equation}
  \hat \mu^{(-m)}_{k}(\{t^*\}) = \frac{\left\lceil t^*\right\rceil - t^*}{\left\lceil t^*\right\rceil - \left\lfloor t^* \right\rfloor}
  \hat \mu^{(-m)}_{k}(\{\left\lceil t^* \right \rceil\}) +
  \frac{t^*-\left\lfloor t^*\right\rfloor}{\left\lceil t^*\right\rceil - \left\lfloor t^* \right\rfloor}
  \hat \mu^{(-m)}_{k}(\{\left\lfloor t^* \right \rfloor\}).
\end{equation}
Interpolation for $\pi$ is done similarly.

With this, we choose the regularization parameters $\lambda_{\mu}$ and
$\lambda_{\pi}$ as the values that minimize the cross validation score
$CV_M(\lambda_\mu, \lambda_\pi)$:
\begin{equation*}
  (\hat \lambda_\mu, \hat \lambda_\pi) = \argmin_{\lambda_\mu \in L_\mu, \lambda_\pi \in L_\pi} CV_M(\lambda_\mu, \lambda_\pi),
\end{equation*}
among pairs of values $(\lambda_\mu, \lambda_\pi)$ in a logarithmically spaced
two-dimensional grid of values $L_\mu \times L_\pi$.

\subsection{Soft gating} \label{subsec:soft_gating}

\begin{figure}[ht!]
  \centering
  \includegraphics[width=.7\textwidth]{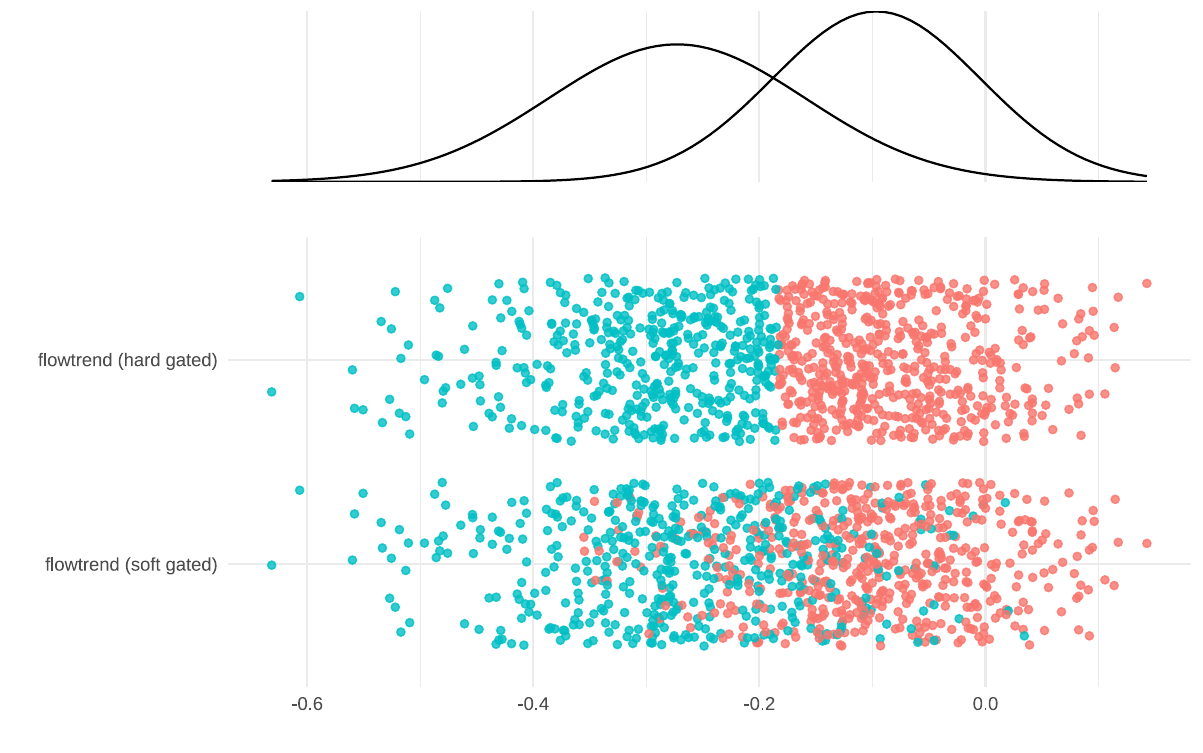}
  \caption{\it Hard versus soft gating. The data are from a particular signal
    size and time point ($\Delta=2$ and $t=60$) of simulated data, described
    shortly in Section~\ref{sec:NumericExperiments}. The $n_t=1,000$ paticles'
    values are heavily overlapping, hence jittered for ease of
    visualization. Displayed on top are the two flowtrend-estimated Gaussian
    density curves of the two clusters. The colors of the points show the drawn
    cluster memberships from the two styles (soft and hard) of gating --
    described in Section~\ref{subsec:soft_gating}. }
  \label{fig:hard-vs-soft-gating}
\end{figure}

When gating is done manually, each particle is assigned to a single cell type.  The responsibilities  from \eqref{eqn:resp_def} offer a probability vector for each particle instead:  
\begin{equation*}
\gamma_{itk}(\hat \mu, \hat \Sigma, \hat \pi) = \frac{\hat \pi_{kt} \cdot  \phi\left(y_i^{(t)}; \hat \mu_{kt\cdot}, \hat \Sigma_k\right)}{\sum_{j = 1}^K \hat \pi_{jt} \cdot  \phi\left(y_i^{(t)}; \hat \mu_{jt\cdot}, \hat \Sigma_j\right)} \text{ for $k=1,\ldots,K$}.
\end{equation*}
If one wishes to assign a particle to a single cell type, we consider two approaches.
A randomized strategy, which we call ``soft gating'' draws from this categorical probability distribution:
$$\hat z^{\text{soft}}_i  = k \text{ with probability } \gamma_{itk}(\hat \mu, \hat \Sigma, \hat
\pi), k= 1,\cdots, K.$$ 
By contrast, ``hard gating'' is deterministic and picks
the particle's membership according to the largest responsibility:
$$\hat z^{\text{hard}}_i = \argmax_{k=1,\cdots,K} \gamma_{itk}(\hat \mu, \hat
\Sigma, \hat \pi).$$ The difference is illustrated in
Figure~\ref{fig:hard-vs-soft-gating}, which shows  soft and hard gating of the
particles from simulated data at a particular time, based on an estimated
{\tt flowtrend} model. In our paper, we opt for soft gating $\hat z_i^{\text{soft}}$ for
gating, since the resulting gated particles resemble the estimated
clusters' distributions more closely.

\section{Numerical experiments}\label{sec:NumericExperiments}

To our knowledge, ours is the first methodology developed for the task of gating a time series of cytograms where the cluster locations and probabilities are both changing across time, without being constrained by environmental covariates. In this section, we demonstrate the performance of our method numerically.
For the sake of comparison, we present two baseline procedures
(and an ``oracle'' model) that take gating procedures designed for single
cytograms and extend them to be applied to a time series of cytograms.

\subsection{Baselines using Gaussian mixtures} \label{subsec:baseline_descrip}

\begin{description}
\item[Oracle model:] The oracle clusters $\{y_i^{(t)}\}_{i,t}$ by generating memberships $Z_i^{(t)}$
  with full knowledge of the underlying data generating model at each time
  point. In other words, for each particle we make a single draw from the categorical distribution, using the true parameters
  $(\mu^*, \Sigma^*, \pi^*)$:
  $$\hat Z^{(t)}_i  = k \text{ with probability } \gamma_{itk}( \mu^*,  \Sigma^*, 
  \pi^*), k= 1,\cdots, K.$$
\item[Overfit model (individual cytogram clustering):] In this model, we
  apply a Gaussian mixture model (GMM) to each cytogram individually. This gives a separate estimate of the three parameters $\mu_{kt\cdot}$,
  $\pi_{kt}$, and $\Sigma_{kt}$, the $k$'th cluster mean, probability and
  variance at every time point $t$.  Since the estimated cluster labels
  $1,\cdots, K$ at any time point are arbitrary, we sequentially permute cluster
  labels to make them consistent across time. Specifically, in comparing a
  Gaussian mixture at time $t$ and another at time $t+1$, we first calculate all
  costs $\{c_{ij}; i,j = 1,\cdots,K\}$ between the $i$-th Gaussian $F_i^{(t)}$
  from one mixture model and $j$-th Gaussian $F_j^{(t+1)}$ of the next time
  point's mixture model as a symmetrized Kullback–Leibler (KL) divergence:
  $$c_{ij} = \frac{1}{2} KL(F^{(t)}_i, F^{(t+1)}_j) + \frac{1}{2} KL(F^{(t+1)}_j, F^{(t)}_i).$$

  Then, we use the Hungarian algorithm \citep{kuhn1955hungarian} which uses
  costs $\{c_{ij}\}_{i,j}$ to find the best cluster assignment of $K$
  Gaussians at time $t$ to those at time $t+1$. 

\item[Underfit model (pooled cytogram clustering):] For this model, we pool
  all the particles into a single cytogram and estimate a single Gaussian
  mixture model to this pooled cytogram.
\end{description}
These overfit and underfit models represent two plausible adaptations of
existing finite mixture models for gating cytograms.  They represent two
opposing approaches for gating a time series of cytograms. The overfit model
learns separately from each time point and uses other time points' data only to
connect cluster memberships. The underfit model ignores temporal variation
when finding the two clusters in the data. We will conduct simulations to
compare our {\tt flowtrend} method to these models.

\subsection{Simulation design}\label{subsec:sim_model}

In order to create realistic synthetic data, we generate from two real cell
populations' mixture model parameters estimated from flow cytometry data in the
Gradients 2 oceanographic research cruise from the Seaflow datasets
\citep{ribalet2019seaflow}. From the estimated
\texttt{flowmix} mixture of experts model on this dataset (see
\cite{hyun2020modeling} for more detail), we isolated our attention to the
cell diameter (the first of the $d=3$ dimensions), and took two clusters representing two
subpopulations of phytoplankton. The first population, according to expert annotation, is a smaller-sized {\it \textit{Prochlorococcus}} picoplankton population, and the second
population is a larger-sized {\it PicoEukaryote} population. The
two populations' mean diameters $\{\mu_{1t1}\}_t$ and $\{\mu_{2t1}\}_t$ and log
odds $\{\alpha_{1t}:=\log (\pi_{1t}/\pi_{2t})\}_t$  were smoothed using
trend-filtering (with degree $l=2$ for means, $l=1$ for $\{\gamma_{1t}\}_t$,
cross-validated), to make them piecewise quadratic and piecewise linear. The smoothed
$\gamma_{1t}$ was further transformed back to probabilities by
$\pi_{1t} = \frac{1}{1+\exp{\gamma_{1t}}}$ and $\pi_{2t} = 1-\pi_{1t}$.

Out of these two smoothed means, we kept the first cluster mean in its same place, and shifted the second cluster mean.  First denote
the mean of the first cluster $\mu_{1\cdot 1}^{\Delta} \in \mathbb R^{T}$ as equal to $\mu_{1 \cdot 1}$ for all $\Delta=0,\cdots, 12$. The
second cluster mean $\mu_{2 \cdot 1}^\Delta$ is then defined by the offset index
$\Delta$ so that the average of $\mu_{2\cdot1}^{\Delta}$ is equal to that of
$\mu_{1 \cdot 1}$ when $\Delta=0$, and at the other end, to be at its
original position in the real estimated model when $\Delta=12$:
$$\mu_{2 \cdot 1}^{\Delta} = \mu_{2 \cdot 1} + \frac{\Delta}{12} \cdot (\bar \mu_{2 \cdot 1} - \bar \mu_{1 \cdot 1}),$$
denoting as $\bar \mu^{(k)} = (1/T) \cdot \sum_{t} \mu_{kt1}, k=1,2$ the averages of
each cluster mean.

For each signal size $\Delta$, we generated $n_t$ particles per time
point, $\{y_{i}^{(t)}, t = 1,\cdots, 296, i=1,\cdots, n_t\}$, from a mixture of
Gaussians, whose latent membership $z_i^{*, (t)}$ was drawn to be $k$ with
probability $\pi_{kt}$. Conditional on membership $k$, the particle $y_i^{(t)}$
was generated as
\begin{equation*}
  \left( y_i^{(t)} \mid {z}_i^{*,(t)} =k \right) \sim \cN(\mu_{\cdot t}^{\Delta},
  \sigma_{k}^2)\text{ for } k = 1,2,
\end{equation*}
where $\sigma_{1} = 0.0918$ and $\sigma_{2} = 0.114$ were estimates taken from
the estimated \texttt{flowmix} model. The resulting data are shown in the top
panel of Figure~\ref{fig:sim}. For a robust simulation, we repeated 10 simulations
for each signal size $\Delta$ and for different sample sizes
$n_t = 100,200,300,400,500$.

\subsection{Simulation results}\label{subsec:sim_results}

\begin{figure}[ht!]
  \centering
  \includegraphics[width=\linewidth]{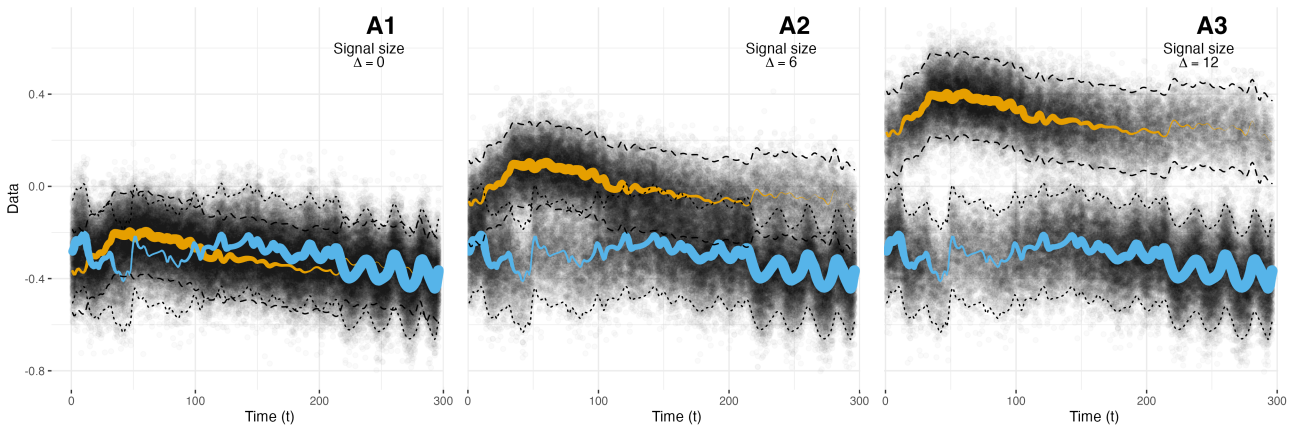}
  \includegraphics[width=.4\linewidth]{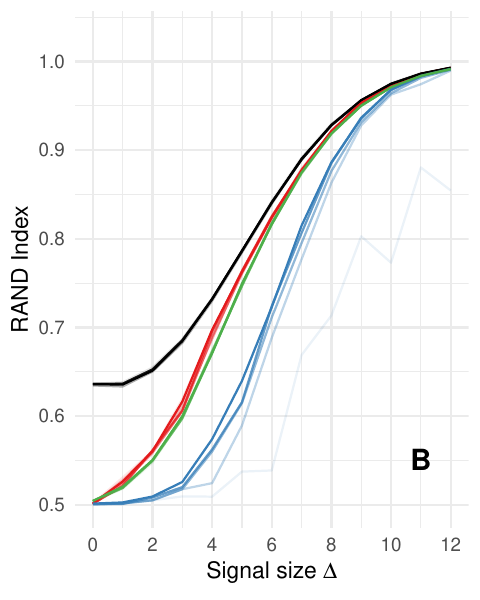}
  \includegraphics[width=.5\linewidth]{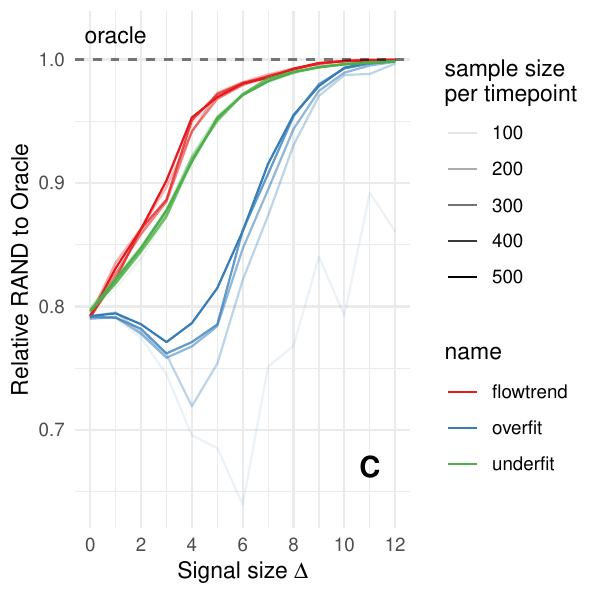}
  \caption{\it Simulation data and results. Panel~A1, A2 and A3 show examples
    of simulated data used in Section~\ref{subsec:sim_results} with signal size
    $\Delta=0,6$ and $11$. In the simulation, the signal size $\Delta$ is varied from $0$
    (where the lower cluster mean $\mu_{1t1}$ and upper cluster mean $\mu_{2t1}$ over $t$ are maximally
    overlapping) to $12$ (corresponding to their original amount of separation in the
    real data). The two curves are smoothed cell diameter means
    originating from actual data, and their thickness represent cluster probabilities, which are also
    smoothed versions from actual data. The dotted and dashed lines show the $\pm 1.96$ standard deviation regions around the cluster means of the two populations. In the bottom panels, the points show performance from 10 simulation repetitions
    at each signal size $\Delta=0,1,\cdots, 12$.  Panel~B
    shows the Rand index of the three models in colors, and the oracle model in
    black. Panel~C shows the relative ratio of the three
    models' Rand index to the oracle's Rand index. }

  \label{fig:sim}
\end{figure}

We begin by examining the overall quality of the estimation by
\texttt{flowtrend} using oceanographic context. The first of the two clusters
corresponds to a {\it \textit{Prochlorococcus}} plankton population, whose true mean
$\mu_{1 \cdot 1}$ has a distinct and repeated daily oscillation between time points 167
through 196 (see Panel A of Figure~\ref{fig:sim}). This oscillation is
due to a clear cell division and photosynthesis pattern driven by the intensity
of sunlight throughout a 24-hour cycle; we would hope for a good model to
capture this. Indeed, our \texttt{flowtrend} model estimates this well -- across
all signal sizes, our method incurs a very small error in model fit.

We measure gating performance using the Rand index
\citep{rand-index},
which compares two partitions of a set of points by calculating the proportion of  the time a pair of points are in the same set in both partitions. That is, we compare an estimated partition $\{\hat z_i^{(t)}\}_t$ to the true partition $\{z_i^{*, (t)}\}_{i,t}$ as
\begin{equation*}
  \text{Rand index} = \frac{\sum_{t_1, t_2 \in \{1,\cdots, T\}}\sum_{1\le i_1 \le n_{t_1}, 1 \le i_2 \le n_{t_2}, i_1\neq i_2 \text{ if } t_1 = t_2} \one(\hat z_{i_1}^{(t_1)} =  z_{i_2}^{*,(t_2)}) }
{ \binom{N}{2}  - \sum_{t=1}^T n_t}
\end{equation*}

We compare the clustering performance of \texttt{flowtrend} to the overfit, underfit, and oracle models described in Section
\ref{subsec:baseline_descrip}.

The bottom row of Figure~\ref{fig:sim} summarizes the simulation results. For
most signal sizes $\Delta$, {\tt flowtrend} outperforms the two baseline models
(Panel B and C of Figure~\ref{fig:sim}). The overfit model gates noticeably
poorly compared to the underfit or {\tt flowtrend} model; this is because the
estimated models are highly varying, and the automatic matching is also poor
when the signal size $\Delta$ is low. The underfit model -- which has constant
means, variance, and probability over $t$ -- seems to have gating performance
closer to that of {\tt flowtrend}, despite having estimated model parameters
very different from the truth. This is because in our simulation example,
precisely estimating the clusters' location and spread in the latter time points
is relatively unimportant for gating performance -- since one of the clusters
has very small probability. But we note that over all values of $\Delta$, (i)
compared to the oracle (Panel~C of Figure~\ref{fig:sim}) {\tt flowtrend} does
significantly better, and (ii) the estimated {\tt flowtrend} model parameters
are very close to the truth (Appendix~A
).

\section{Application to Seaflow data}
\label{sec:3dreal}

We apply our \texttt{flowtrend} model to the Gradients 2 cruise from the Seaflow
datasets \citep{ribalet2019seaflow}. The cytograms from this
cruise are recorded almost continuously, but aggregated to each unique hour for
a total of $T=296$ time points, and consist of $d = 3$ optical properties of the
cells, as explained earlier in the paper. All original measurements (cell
diameter, red fluorescence, and orange fluorescence) have been log transformed,
and shifted/scaled to be between about 0 and 8.

\begin{figure}[ht!]
  \includegraphics[width=\linewidth]{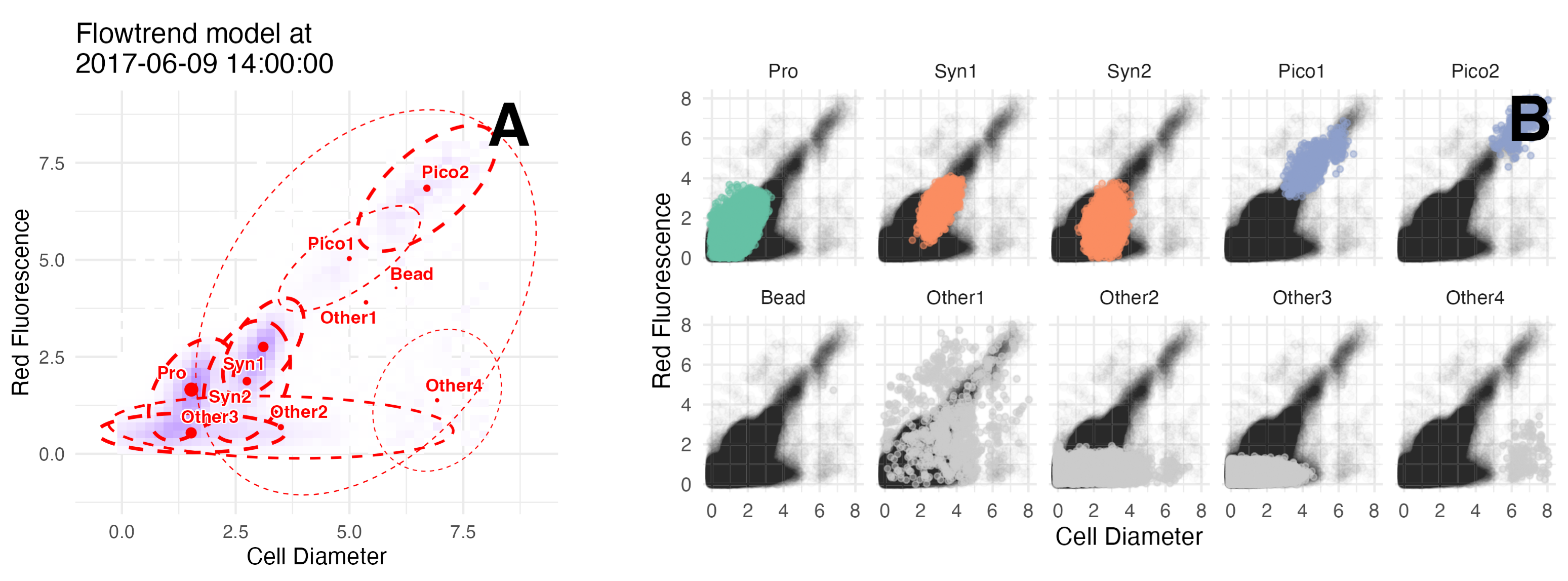}
  \includegraphics[width=\linewidth, trim={0 0 0 .9cm},clip]{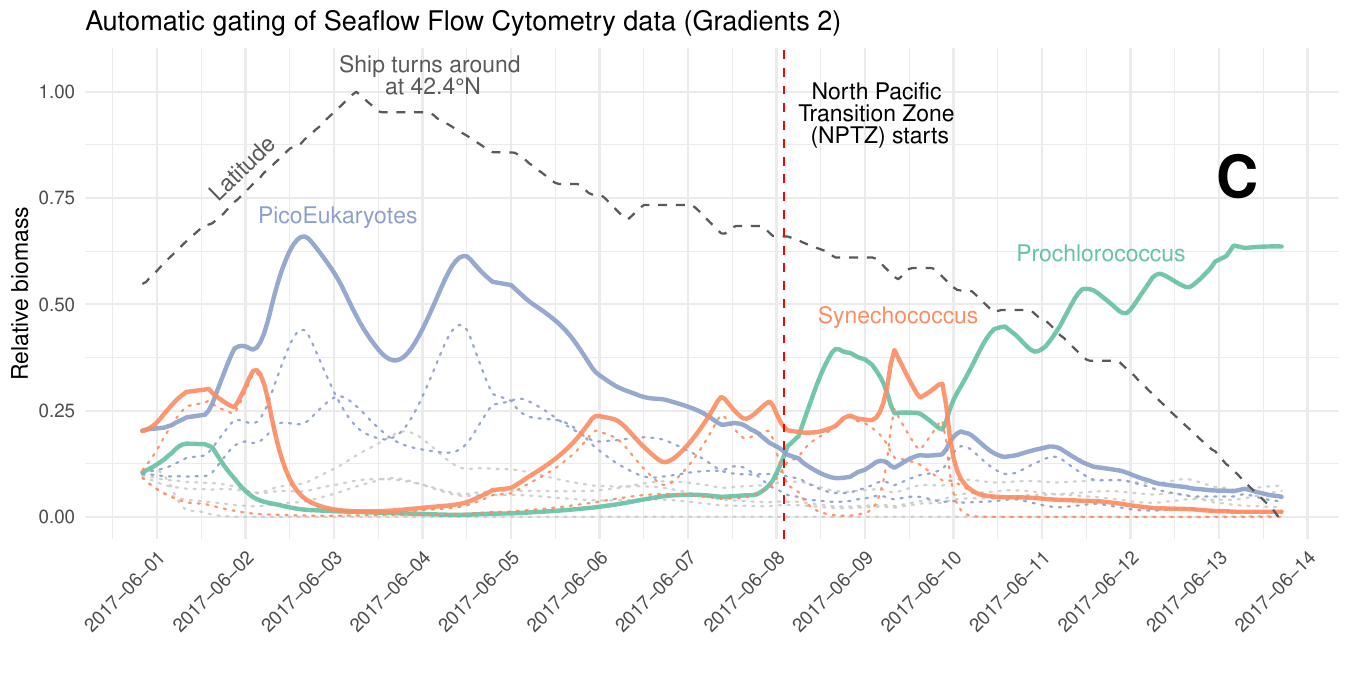}
  \caption{\it [Top row, panels A \& B] The top row shows the estimated
    10-cluster flowtrend model (Panel A) used for probabilistically classifying
    particles (Panel B). Out of 296 time points, only one time point
    (2017-06-09, at 14:00:00 UTC) is shown in these panels. Each particle is
    soft-gated, as described in Section~\ref{subsec:soft_gating} In Panel A, the
    estimated 95\% probability ellipses are shown in dashed red lines, and the
    solid red circles represent the cluster means -- all plotted over a backdrop
    showing the binned particles' distribution. In Panel B, all particles are
    shown in each of the ten subpanels, in transparent black points. The solid
    colored points mark the gated particles assigned to each cluster. The
    cluster labels on the panel titles are from expert annotation, and stand for
    major picoplankton species such as \textit{Prochlorococcus} (Pro), \textit{Synechococcus}
    (Syn1, Syn2), PicoEukaryote (Pico1, Pico2).  [Bottom row, Panel~C] The
    estimated relative biomass over time are shown in Panel~C. Thick solid
    colored lines show the total relative abundance for \textit{Prochlorococcus},
    \textit{Synechococcus} and PicoEukaryotes. The latter two subspecies each have two
    associated clusters identified by our model, whose relative abundance over
    time is shown in thinner dotted lines of the same color. (Grey dashed lines
    show the relative abundance of estimated clusters that do not clearly match
    with a major species).
  }
  \label{fig:gated}
\end{figure}

We first perform cross validation to select tuning parameters for the \texttt{flowtrend} model, 
resulting in parameter
estimates $\hat \mu, \hat \Sigma, \hat \pi$. Using these, we soft-gated the particles as described in Section~\ref{subsec:soft_gating}. Figure~\ref{fig:gated} shows the resulting gated subpopulations of picoplankton.  The cluster names come from having an expert look at the gates and annotate. The labels stand for picoplankton
species such as \textit{Prochlorococcus}, \textit{Synechococcus}, and {\it PicoEukaryotes}. 
Our method successfully identified one subpopulation as a calibration bead, which does not change over time in cytogram space.

Panel~C of Figure~\ref{fig:gated} shows the relative biomass of the
estimated {\tt flowtrend} model clusters over time  (represented by $\hat \pi_{kt}$ for cluster $k$ at time $t$). The dashed lines show the
original ten $\pi$ estimates in colors according to the expert-annotated
microbial species. Out of the ten clusters, we will focus our attention on the
clusters that correspond to known species of picophytoplankton. The thick solid
lines show the sum of certain clusters corresponding to the same species -- for
instance, \textit{Synechococcus} was detected as two different clusters (labeled Syn1 and
Syn2 in Figure~\ref{fig:gated}), and the sum of their $\pi_{kt}$ at each time
point represents the total estimate of \textit{Synechococcus} over time. We can see a
gradual increase of \textit{Prochlorococcus} biomass as the cruise passes the
North-Pacific transition zone.  We also see a thriving of PicoEukaryotes
relative to \textit{Synechococcus} and \textit{Prochlorococcus} in northern, cold and
nutrient-rich waters. Indeed, along the north-south transect line, there are
different ``niches'' where each of these three major subpopulations seem to
thrive.

\begin{figure}
  \includegraphics[width=\linewidth]{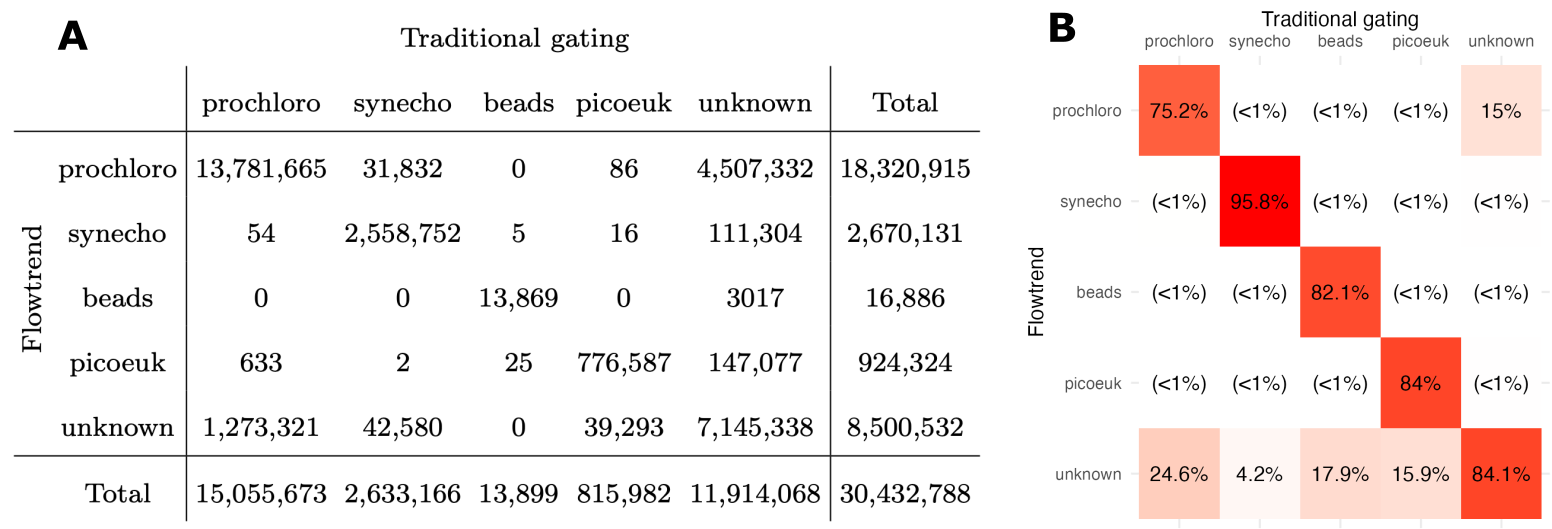} 
  \caption{\it [Panel A] A five by five contingency table comparing
    {\tt flowtrend} gating and traditional gating. Each gating method identified five
    major subpopulations across over 30 million particles collected on the Gradients2 cruise. The overall
    agreement is high (Rand Index $0.7280$). [Panel B] Each row of the contingency table is scaled to sum to 100\% and shown as a heatmap. For instance, the first row shows the estimated
    relative class frequency from traditional gating, out of the the 18,320,915
    cells that {\tt flowtrend} found to be \textit{Prochlorococcus}. 91.54\% of those
    cells are also found to be \textit{Prochlorococcus} by traditional gating. }
\label{tab:flowtrend-vs-annette}

\end{figure}

\begin{figure}[ht!]
  \centering
  \includegraphics[width=.8\linewidth, clip, trim={0 0 0 20}]{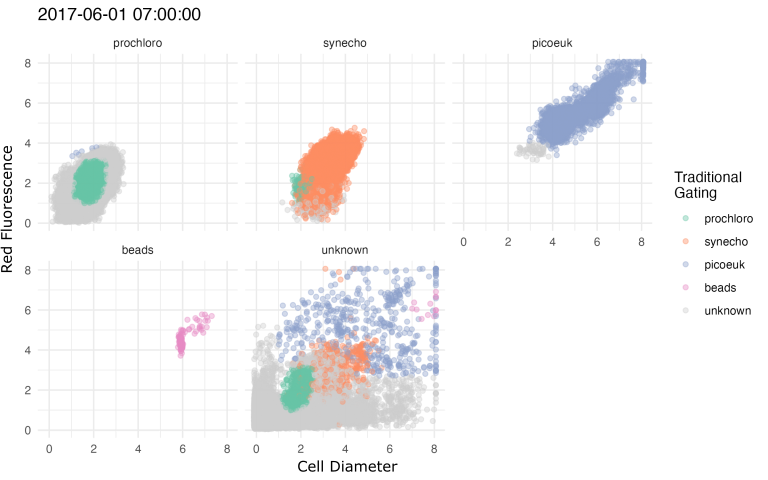}
  \caption{\it Gating comparison at one time point (2017-06-01, during the hour
    starting at 07:00:00 UTC).  The same data was clustered and expert-annotated
    in two ways -- using our {\tt flowtrend} method, and using traditional
    gating (described in \citealt{ribalet2019seaflow}). In each of the five panels,
    we show the gated major cell subpopulations of interest (\textit{Prochlorococcus},
    \textit{Synechococcus}, Beads, and PicoEukaryotes) by {\tt flowtrend}. In those same
    plots, the color of the points show the gated particles using traditional
    gating. }
\label{fig:flowtrend-vs-annette}
\end{figure}

Next, we compare the gating results from {\tt flowtrend} to what we will refer to as
{\it traditional gating}.  From the manuscript \cite{ribalet2019seaflow} that
originally processed and analyzed this data, the authors' gating strategy is
written as follows: ``The classification of particles into cell populations was
conducted uniformly across all samples using a combination of manual gating and
unsupervised clustering algorithms,'' such as \texttt{flowDensity}
\citep{flowdensity}.  Table~\ref{tab:flowtrend-vs-annette} shows the cluster
expert-annotated membership comparison, and
Figure~\ref{fig:flowtrend-vs-annette} visually compares all particles' gating results at one
time point. 
We now make some observations based on the results.
It is visually clear that the traditional gating employs hard-gating while {\tt
  flowtrend} uses soft-gating (as described in
Section~\ref{subsec:soft_gating}). Nonetheless, the overall agreement of the two
clusterings is quite high. The bottom panel of
Table~\ref{tab:flowtrend-vs-annette} shows the row-normalized
contingency table as a heatmap, showing that almost all ($>90\%$) particles
gated as a known major cell subspecies (abbreviated as prochloro, synecho,
beads, or picoeuk) by {\tt flowtrend} were labeled the same by traditional
gating. The Rand index of all particles across all time points is about
$0.7280$.

If we examine the Rand
index at each time point separately, we can see that the Rand is quite high
before crossing the transition zone, and becomes abruptly lower after it (see
Appendix~A
).
Fewer than one-tenth of \textit{Prochlorococcus} particles according to traditional
gating are classified as unknown by {\tt flowtrend}. This points to a mistake
made by traditional gating on data directly after crossing the transition zone
from north to south. However, 33\% of particles that {\tt flowtrend} deems to be
\textit{Prochlorococcus}, traditional gating calls unknown. This is caused by the
traditional gating's hard-gating strategy failing to capture \textit{Prochlorococcus}
particles that overlap with the unknown (debris) cluster. The two methods easily
agree on PicoEukaryotes, since they are always located in the top-right corner
of diameter-chlorophyll plots (in Figure~\ref{fig:flowtrend-vs-annette}).  The
agreement in \textit{Synechococcus} is also strikingly high. Both agree on most
($>95\%$) particles. This is due to \textit{Synechococcus} being easier than \textit{Prochlorococcus} to gate since it is clearly separable
in orange fluorescence.  The
agreement on beads is also relatively high.

\section{Discussion}

In this paper, we propose a new trend-filtered mixture-of-experts model that
specializes in automatic gating flow cytometry data continuously collected over
time.  We conducted a realistic synthetic simulation as well as a careful
application to real ocean flow cytometry data collected from the North Pacific
ocean. Our method gates the cells in this dataset very similarly to a human-intensive, 
hand-gated approach from a previous study.  Doing so reveals additional insights
about the niches in the North Pacific ocean observed in terms of relative
abundances of major microbial species.  Since there is no suitable gating method in the
literature a time-series of marine flow cytometry, an automatic tool such as ours will be a valuable addition to the literature.

The idea of fitting a mixture of trend filtering models is, to our knowledge, entirely novel. A problem-specific feature of our approach is the explicit constraint on the
cluster means, which comes from biological understanding of phytoplankton.

Estimating the model properly requires many repeated applications of the
algorithm to different slices of the data and hyperparameters. The computational
cost is quite large for users without access to a parallel computing system or
many CPU cores. It will be important in future work to develop a model seletion
strategy (e.g. information criteria) to reduce the overall computational cost.
Also, statistical inference for model parameters such as relative abundance
$\pi$ and membership probabilities $P(z_{it}=k\mid y_i^{(t)})$ 
would be valuable for
scientists.

\section{Acknowledgments}

This work was supported by grants by the Simons Collaboration on Computational Biogeochemical Modeling of Marine Ecosystems/CBIOMES (Grant ID: 549939 to JB, Microbial Oceanography Project Award ID 574495 to FR, 1195553 to SH). We also thank Chris Berthiaume and Dr. Annette Hynes for their help in processing and curating SeaFlow data. The
authors also acknowledge the Center for Advanced Research Computing (CARC) at the University of Southern California for providing computing resources that have contributed to the research results reported within this publication. URL: https://carc.usc.edu.

%% file: supp-content.tex
\section*{Appendix A: Additional figures}
\label{sec:appendix-additional-figures}

\begin{figure}[ht!]
  \centering
  \includegraphics[width=.7\linewidth]{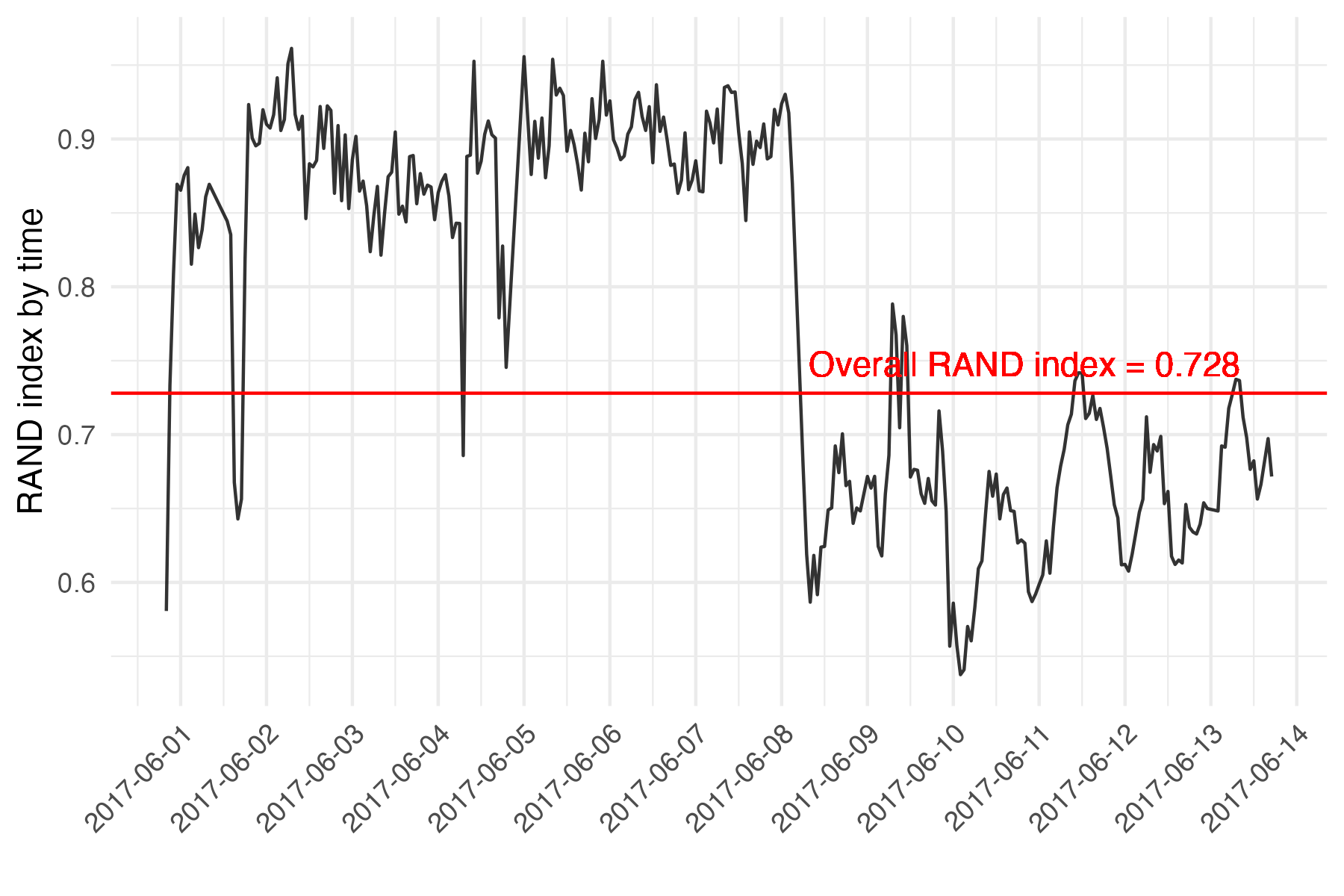}
  \caption{\it Rand index calculated at each time point between flowtrend
    soft-gated memberships and traditional gating memberships, on real flow
    cytometry data described in
    Section~4.
    There is a high agreement
    around $0.9$ before the transition zone crossing at about time point
    $t=170$, and lower agreement afterwards.}
  \label{fig:rand-over-time}
\end{figure}

Figure~\ref{fig:rand-over-time} shows the Rand index calculated separately for cytograms at each time point.

\begin{figure}[ht!]
  \centering
  \includegraphics[width=1\linewidth]{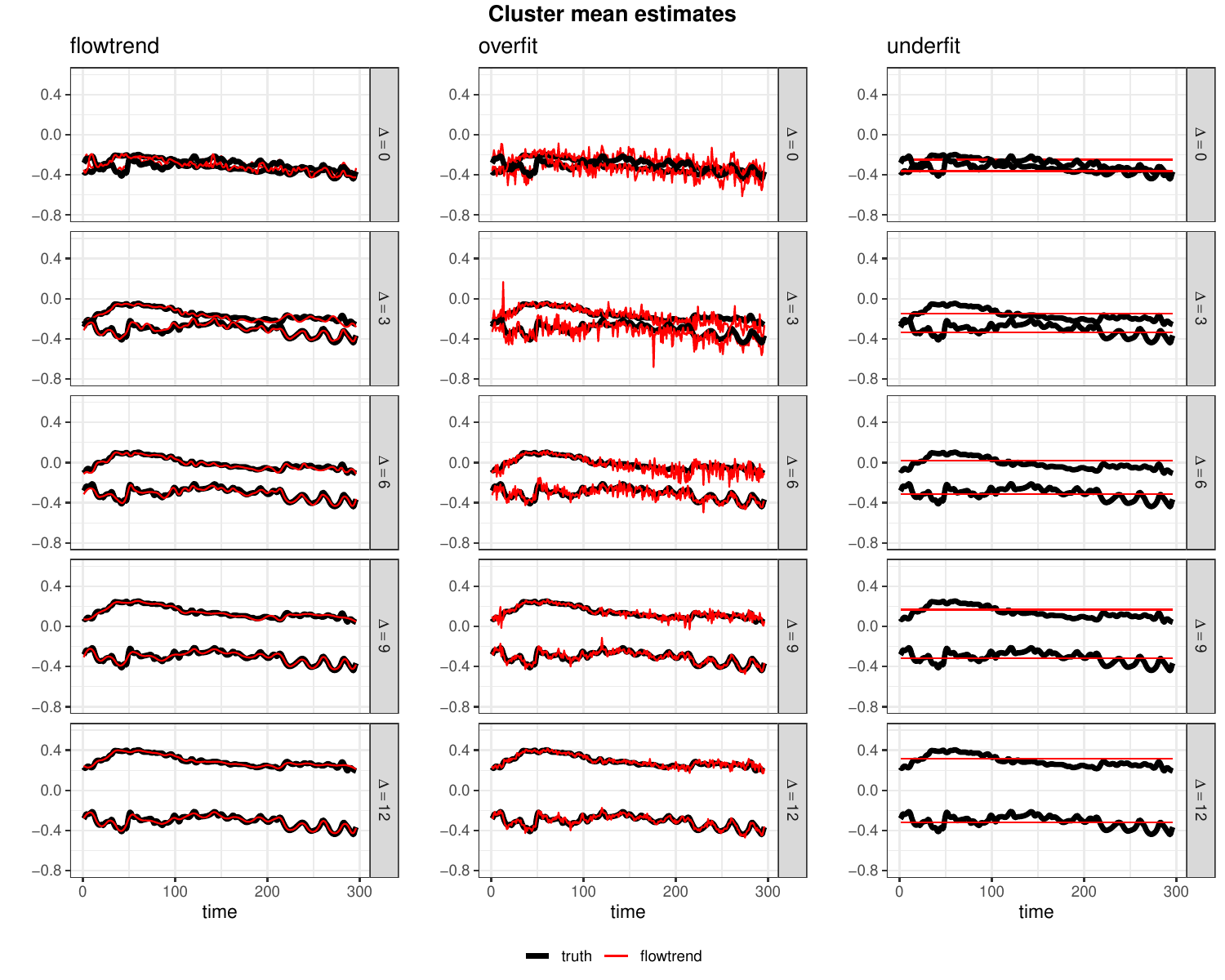}
  \caption{\it Cluster mean estimates from simulated datasets described in
    Section~3
    are shown here. Each column of panels
    shows the estimates from the same model, over several signal sizes. The mean
    estimates by our proposed flowtrend model are the most accurate, over all
    signal sizes.}
  \label{fig:fits1}
\end{figure}

\begin{figure}[ht!]
  \centering
  \includegraphics[width=1\linewidth]{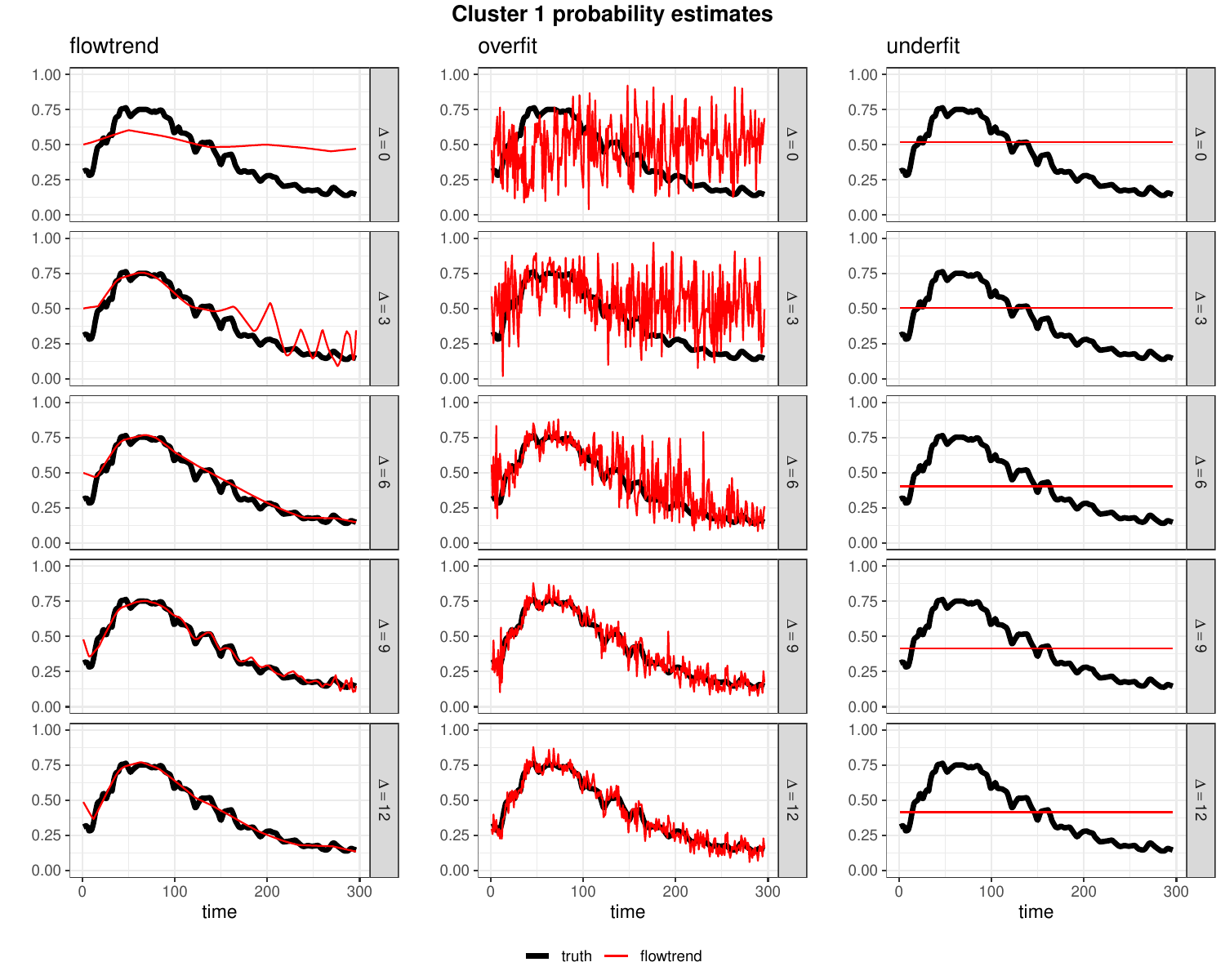}
  \caption{\it The probability estimates for one of the two clusters from
    simulated datasets described in
    Section~3
    are
    shown here. Each column of panels shows the estimates from the same model,
    over several signal sizes. The other cluster's estimates are omitted because
    they are redundant. The probability estimates by our proposed flowtrend
    model are the most accurate, over all signal sizes.}
  \label{fig:fits2}
\end{figure}

Figure~\ref{fig:fits1} and \ref{fig:fits2} show the fitted mean and probabilities for the
simulation example, at several signal sizes.

\newpage

\newpage

\newpage
\section*{Appendix B: ADMM Details}
\label{sec:appendix-admm-details}

For a specific cluster $k$, the ADMM algorithm minimizes over each of the variables
 $\mu_{k \cdot \cdot}, w, z, \{u_w^{(j)}\}_1^d$, and
 $ \{u_z^{(t)}\}_1^T $ in the augmented Lagrangian:
\begin{align*}
  L(\mu_{k\cdot\cdot}, w,  z, \{u_w^{(j)}\}_1^d,
  \{u_z^{(t)}\}_1^T)
  =& 
     \frac{1}{2N} \sum_{t=1}^T
     \sum_{i=1}^{n_t} \tilde{\gamma}_{itk} (y_i^{(t)} - \mu_{k t \cdot})^\top \hat{\Sigma}_k^{-1} ( y_i^{(t)} - \mu_{k t \cdot}) \\
   &+ \lambda_\mu \sum_{j=1}^d \|D^{(1)}w_{\cdot j}\|_1 + \sum_{t=1}^T \one_\infty(\|z_{t\cdot}\|_2 \le r)\\
   &+ \sum_{t=1}^T \left[ {u_z^{(t)}}^\top ( \mu_{k t \cdot} - \bar{\mu}^{(k)} -z_{t \cdot} )
     + \frac{\rho}{2} \|\mu_{k t \cdot} - \bar{\mu}^{(k)} - z_{t \cdot}  \|^2 \right]\\
   & + \sum_{j=1}^d \left[ u_w^{(j)^\top}(D^{(l_\mu)}\mu_{k \cdot j} - w_{\cdot j} ) + \frac{\rho}{2} \|D^{(l_\mu)}\mu_{k \cdot j} - w_{\cdot j} \|^2 \right].
\end{align*}
We provide details of the ADMM algorithm next.
\begin{enumerate}

\item[1.] $\mu_{k\cdot\cdot} \leftarrow \argmin_{\mu_{k \cdot \cdot}}
  L(\mu_{k\cdot\cdot}, w,  z, \{u_w^{(j)}\}_1^d, \{u_z^{(t)}\}_1^T)$

  To start, let $\eta$ denote the $d \times T$ matrix $\mu_{k \cdot\cdot}^T$ for
  cluster $k$. Then, we will rewrite several of the quantities in the augmented
  Lagrangian $\eta$. Let $e_c$ denote a vector that has 1 in the $c$-th entry
  and 0 elsewhere. Then we write $\mu_{kt \cdot}$ as
  $\eta e_t \in \mathbb R^{d \times 1}$, the time-wise average $\bar{\mu}^{(k)}$
  as $\frac{1}{T}\eta \one_T \in \mathbb R^{d \times 1}$, and
  $\mu_{k\cdot j}^\top$ as $e_j^\top \eta \in \mathbb R^{T \times 1}$,
    so that our minimization with respect to $\eta$ is:

    \begin{align}
      \eta \leftarrow \argmin_\eta \;\;
      & \frac{1}{2N} \sum_{t=1}^T
        \sum_{i=1}^{n_t} \tilde{\gamma}_{itk} (y_i^{(t)} - \eta e_t)^\top \hat{\Sigma}_k^{-1} ( y_i^{(t)} - \eta e_t) \nonumber\\
      & + \sum_{t=1}^T \left[ {u_z^{(t)}}^\top ( \eta e_t - \frac{1}{T}\eta \one_T  -z_{t \cdot} )
        + \frac{\rho}{2} \|\eta e_t - \frac{1}{T}\eta \one_T  - z_{t \cdot}  \|^2 \right] \nonumber\\
      & + \sum_{j=1}^d \left[ u_w^{(j)^\top}(D^{(l_\mu)}\eta^\top e_j - w_{\cdot j} ) + \frac{\rho}{2} \|D^{(l_\mu)}\eta^\top e_j - w_{\cdot j} \|^2 \right].
        \label{eq:abc}
    \end{align}
    Now, we calculate the derivative of the objective in \eqref{eq:abc} with
    respect to $\eta$. We will examine one term at a time. The derivative of the
    first term is:
    \begin{align*}
      \frac{\partial}{\partial \eta} \left[\frac{1}{2N} \sum_{t=1}^T
      \sum_{i=1}^{n_t} \tilde{\gamma}_{itk} (y_i^{(t)} - \eta e_t)^\top \hat{\Sigma}_k^{-1} ( y_i^{(t)} - \eta e_t) \right] &= \frac{1}{2N} \sum_{t=1}^T
                                                                                                                       \sum_{i=1}^{n_t} \tilde{\gamma}_{itk}  \frac{\partial}{\partial \eta} (y_i^{(t)} - \eta e_t)^\top \hat{\Sigma}_k^{-1} ( y_i^{(t)} - \eta e_t) \\
                                                                                                                     &= \frac{1}{2N} \sum_{t=1}^T
                                                                                                                       \sum_{i=1}^{n_t} -2 \tilde{\gamma}_{itk} \hat{\Sigma}_k^{-1} (y_i^{(t)} - \eta e_t)e_t^\top.
    \end{align*}
    The derivative of the second term is:
    \begin{align*}
      \frac{\partial}{\partial \eta}&\sum_{t=1}^T \left[ {u_z^{(t)}}^\top ( \eta e_t - \frac{1}{T}\eta \one_T  - z_{t \cdot} )
      +  \frac{\rho}{2} \|\eta e_t - \frac{1}{T}\eta \one_T  - z_{t \cdot}  \|^2 \right]  \\
      &= \sum_{t=1}^T \frac{\partial}{\partial \eta}\left[ {u_z^{(t)}}^\top ( \eta [e_t - \frac{1}{T}\one_T] -z_{t \cdot} )
      + \frac{\rho}{2} \|\eta [e_t - \frac{1}{T}\one_T]  - z_{t \cdot}  \|^2 \right] \\
      &= 
      \sum_{t=1}^T \left[ {u_z^{(t)}} [e_t - \frac{1}{T}\one_T]^\top
      - \rho [\eta [e_t - \frac{1}{T}\one_T]  - z_{t \cdot}  ][e_t - \frac{1}{T}\one_T]^\top\right].
    \end{align*}
    Lastly, the third term can be rearranged using:

    \begin{equation*}
      u_w^{(j)^\top}(D^{(l_\mu)}\eta^\top e_j - w_{\cdot j} ) = e_j^\top \eta (D^{(l_\mu)})^\top u^{(j)}_w - w_{\cdot j}^\top u_w^{(j)},
    \end{equation*}
    and
    \begin{equation*}
      \|D^{(l_\mu)}\eta^\top e_j - w_{\cdot j} \|^2_2 =
      \text{Trace}\left[ \left( e_j^\top \eta (D^{(l_\mu)})^\top - w_{\cdot j}^\top\right)\left( e_j^\top \eta (D^{(l_\mu)})^\top - w_{\cdot j}^\top\right)^\top \right]. 
    \end{equation*}
    which uses that $\| x \|_2^2 = \| x^\top \|_F= \text{Trace}(xx^\top)$. So,
    the derivative of the third term is:
    \begin{align*}
      \frac{\partial}{\partial\eta}\sum_{j=1}^d&\left[ u_w^{(j)^\top}(D^{(l_\mu)}\eta^\top e_j - w_{\cdot j} ) + \frac{\rho}{2} \|D^{(l_\mu)}\eta^\top e_j - w_{\cdot j} \|^2 \right] \\
                                              &=  \frac{\partial}{\partial\eta}\sum_{j=1}^d\left[ e_j^\top \eta (D^{(l_\mu)})^\top u^{(j)}_w - w_{\cdot j}^\top u_w^{(j)} + \frac{\rho}{2} \|e_j^\top \eta (D^{(l_\mu)})^\top - w_{\cdot j}^\top\|^2 \right] \\
                                              &= \sum_{j=1}^d\left[e_j(u_w^{(j)})^\top D^{(l_\mu)} + \rho e_j\left(e_j^\top \eta (D^{(l_\mu)})^\top - w_{\cdot j}^\top \right) D^{(l_\mu)}\right].
    \end{align*}
    Combining these, the overall derivative of the augmented Lagrangian with
    respect to $\eta$ is:
    \begin{equation}
      \begin{split}
        -\frac{1}{N} &\sum_{t=1}^T
          \sum_{i=1}^{n_t} \tilde{\gamma}_{itk} \hat{\Sigma}_k^{-1} (y_i^{(t)} - \eta e_t)e_t^\top  \\
        &+    \sum_{t=1}^T \left[ {u_z^{(t)}} [e_t - \frac{1}{T}\one_T]^\top
          - \rho [\eta [e_t - \frac{1}{T}\one_T]  - z_{t \cdot}  ][e_t - \frac{1}{T}\one_T]^\top\right] \\
        &+ \sum_{j=1}^d\left[e_j(u_w^{(j)})^\top D^{(l_\mu)} + \rho e_j\left(e_j^\top \eta (D^{(l_\mu)})^\top - w_{\cdot j}^\top \right) D^{(l_\mu)}\right].
      \end{split} \label{eqn:mu_diff}
    \end{equation}

    We set this derivative \eqref{eqn:mu_diff} equal to zero to get a matrix
    equation of the form:
    \begin{equation*}
      A\eta B + \eta C + E = 0,
    \end{equation*}
    for matrices $A, B, C$ and $E$ defined as:
      \begin{align*}
        A &= \frac{1}{N} \hat{\Sigma}_k^{-1} \\
        B &=  \text{diag}(\{\tilde{\gamma}_{tk}\}_{t=1}^T)\\
        C &= \rho\left[(D^{(l_\mu)})^\top D^{(l_\mu)} + \sum_{t=1}^T \tilde{e}_t \tilde{e}_t^\top \right] \\
        E &=      \frac{1}{N}\sum_{t=1}^T \sum_{i=1}^{n_t} \tilde{\gamma}_{itk} \hat{\Sigma}_k^{-1}y_i^{(t)} e_t^\top  - \sum_{t=1}^T (u_z^{(t)} - \rho z_{t \cdot}) \tilde{e}_t^\top - \sum_{j=1}^d \left[ e_j(u_w^{(j)})^\top D^{(l_\mu)} - \rho e_jw_{\cdot j}^\top D^{(l_\mu)} \right].
    \end{align*}

    Each term's derivation is explained next. Matrix $E$ is formed from
    the terms in \eqref{eqn:mu_diff} with no $\eta$:
    \begin{equation}
      -\frac{1}{N}\sum_{t=1}^T \sum_{i=1}^{n_t} \tilde{\gamma}_{itk} \hat{\Sigma}_k^{-1}y_i^{(t)} e_t^\top  + \sum_{t=1}^T (u_z^{(t)} - \rho z_{t \cdot}) {\tilde{e}}_t^\top + \sum_{j=1}^d \left[ e_j(u_w^{(j)})^\top D^{(l_\mu)} - \rho e_j w_{\cdot j}^\top D^{(l_\mu)} \right].
      \label{eqn:nomu}
    \end{equation}
    where we have written
    $\tilde{e}_t^\top := \left[e_t - \frac{1}{T}\one_T \right]^\top$.  Next,
    matrix $C$ is formed by collecting the following two terms from
    \eqref{eqn:mu_diff}:
    \begin{equation}
      \eta \cdot \rho \sum_{t=1}^T \tilde{e}_t \tilde{e}_t^\top
      + \sum_{j=1}^d \rho e_j e_j^T \eta \rho (D^{(l_\mu)})^\top D^{(l_\mu)},
      \label{eqn:muright2}
    \end{equation}
    and simplifying to
    \begin{equation*}
      \eta \cdot \left(\rho \sum_{t=1}^T \tilde{e}_t \tilde{e}_t^\top +  \rho
        (D^{(l_\mu)})^\top D^{(l_\mu)}\right) =: \eta C.
    \end{equation*}
    Lastly, $A$ and $B$ are formed by gathering the remaining term:
    \begin{equation*}
      \frac{1}{N}  \hat{\Sigma}_k^{-1}\eta \sum_{t=1}^T
      \sum_{i=1}^{n_t} \tilde{\gamma}_{itk} e_te_t^\top ,
    \end{equation*}
    which simplifies using $\tilde{\gamma}_{tk} = \sum_i \tilde{\gamma}_{itk}$ to
    \begin{equation*}
      \frac{1}{N}  \hat{\Sigma}_k^{-1}\cdot \eta \cdot \text{diag}(\{\tilde{\gamma}_{tk}\}_{t=1}^T)
      =: A \eta B.
    \end{equation*}
   Notice that $B^{-1}$ can be right-multiplied to get:
\begin{equation} \label{eqn:sylv}
A\cdot\eta + \eta \cdot C B^{-1} + E B^{-1} = 0,
\end{equation}
where $B^{-1} = \diag(\{ \frac{1}{\tilde \gamma_{tk}}) \}_{t=1}^T)$.  This
rearrangement to derive the Sylvester equation in \eqref{eqn:sylv} pays off
since we can now solve it using a fast solver. One such solver can be found in
the \texttt{LAPACK} library in \texttt{C++}. Furthermore, we can take advantage
of that the matrices $A$ and $C B^{-1}$ do not change across ADMM iterations, so
that they can be Schur-decomposed and solved using an even faster Sylvester
equation solver (e.g. Bartels-Stewart) that only uses triangular matrices as
coefficients to a Sylvester equation.

The solution of \eqref{eqn:sylv}, transposed, is used to update $\mu_{k \cdot \cdot}$.

  \item[2a.] $z_{t \cdot} \leftarrow \argmin_{z_{t \cdot}}   L(\mu_{k \cdot \cdot}, w, z, \{u_w^{(j)}\}_1^d, \{u_z^{(t)}\}_1^T)$ for $t=1,\cdots, T$
 
    This update is similar to the update from \citet{hyun2020modeling}, namely
    projecting the current parameter estimates and the auxillary variables onto
    the ball of radius $r$. Specifically,
    \[
      z_{t, \cdot} \leftarrow
      \begin{cases}
        \mu_{kt \cdot} - \bar{\mu}^{(k)} + u_{z}^{(t)}/\rho & \text{if} \ \|\mu_{kt \cdot} - \bar{\mu}^{(k)} + u_{z}^{(t)}/\rho \|_2 \le r \\
        r \frac{ \mu_{k t \cdot} - \bar{\mu}^{(k)} +
        u_{z}^{(t)}/\rho}{\| \mu_{k t \cdot} -
        \bar{\mu}^{(k)} + u_{z}^{(t)}/\rho\|_2} & \text{otherwise}
      \end{cases}.
    \]
  \item[2b.]
    $w_{\cdot j} \leftarrow \argmin_{w_{\cdot j}} L(\mu_{k \cdot \cdot}, w, z, \{u_w^{(j)}\}_1^d, \{u_z^{(t)}\}_1^T) $ for $j=1,\cdots, d$
    
    This update is inspired by \citet{ramdas2016fast}. Take the minimization
    with respect to $w_{\cdot j}$:
    \begin{equation}\label{eq:w-update}
    w_{\cdot j} \leftarrow \argmin_{w_{\cdot j}} \sum_{i=1}^d \left[ u_w^{(i)^\top}(D^{(l_\mu)}\mu_{k \cdot i} - w_i ) + \frac{\rho}{2} \|D^{(l_\mu)}\mu_{i \cdot} - w_i \|^2 + \lambda_\mu \|D^{(1)} w_i\|_1 \right]. 
    \end{equation}
    This can be re-written as a \textit{scaled} problem by completing the
    square. In particular, let $\epsilon_j = D^{(l_\mu)} \mu_{k \cdot j} -
    w_{\cdot j}$. The objective in \eqref{eq:w-update} can be written as:

    \begin{align*}
      u_w^{(i)^\top} \epsilon_i + \frac{\rho}{2} \|  \epsilon_i\|^2 + \lambda_\mu
      \|D^{(1)} w_i\|_1
      &=         u_w^{(i)^\top} \epsilon_i +  \| \sqrt{\rho/2} \epsilon_i\|^2 + \lambda_\mu \|D^{(1)} w_i\|_1 \\
      &= \frac{2}{2\sqrt{\rho/2}} u_w^{(i)^\top}\sqrt{\rho/2} \epsilon_i +  \| \sqrt{\rho/2} \epsilon_i\|^2 + \lambda_\mu \|D^{(1)} w_i\|_1 \\
      &= \frac{2}{\sqrt{2\rho}} u_w^{(i)^\top}\sqrt{\rho/2} \epsilon_i +  \| \sqrt{\rho/2} \epsilon_i\|^2 + \lambda_\mu \|D^{(1)} w_i\|_1 \\
      &= \left\|\sqrt{\rho/2}\epsilon_i +  \frac{1}{\sqrt{2\rho}} u_w^{(i)} \right\|^2 + \lambda_\mu \|D^{(1)} w_i\|_1 -  \frac{1}{2\rho}\| u_w^{(i)^\top} \|^2
    \end{align*}
    leading to the following minimization, after substituting back in $\epsilon$:
    \[
    w_{\cdot j} \leftarrow \argmin_{w_{\cdot j}} \sum_{i=1}^d \left[ \frac{\rho}{2} \|D^{(l_\mu)}\mu_{k \cdot i} - w_i + \frac{1}{\rho}  u_w^{(i)^\top}\|^2 - \frac{1}{2\rho}\| u_w^{(i)^\top} \|^2 + \lambda_\mu \|D^{(1)} w_i\|_1 \right]. 
    \]
    Introducing the pseudo-response $\xi_j = D^{(l_\mu)}\mu_{k \cdot j} + \frac{1}{\rho} u_w^{(j)} \in \R^T$, we see that the above is equivalent to the problem
    \[
    w_{\cdot j} \leftarrow \argmin_{w_{\cdot j}} \sum_{i=1}^d \left[ \frac{\rho}{2} \|\xi_i - w_i \|^2 + \lambda_\mu \| D^{(1)}w_i \|_1\right].
    \]
    Note that each $w_{\cdot j}$ can be optimized separately, so the above is equivalent
    to fitting $d$ fused-lasso problems, for which there exist efficient dynamic
    programming algorithms, such as the algorithm proposed in
    \citet{johnson2013dynamic}. In practice, this means we solve $d$ versions of
    the following problem:
    \[
      \min_{w_{\cdot j}} \frac{\rho}{2} \|\xi_j - w_{\cdot j} \|^2 + \lambda_\mu \| D^{(1)}w_{\cdot j} \|_1
      \iff \min_{w_{\cdot j}} \|\xi_j - w_{\cdot j} \|^2 + \frac{2\lambda}{\rho} \| D^{(1)}w_{\cdot j}
      \|_1.
  \]

 \item[3a.] $u_z^{(t)} \leftarrow u_z^{(t)} + \rho ( \mu_{kt \cdot} - \bar{\mu}^{(k)}- z_{t \cdot} )$ for
   $t=1,\cdots, T$ 
 \item[3b.] $u_w^{(j)} \leftarrow u_w^{(j)} + \rho (D^{(l_\mu)}\mu_{k \cdot j} - w_{\cdot j})$ for $j= 1, \cdots, d$.
    \end{enumerate}
 Steps 1 through 3b are repeated until the objective value numerically stabilizes.

\section*{Appendix C: Lemma 2 from \citet{tibshirani2014adaptive}}
\label{sec:appendix-lemma-tf}

Here, we reproduce Lemma 2 in Section 3.3 from a manuscript on trend filtering
\citep{tibshirani2014adaptive}, where the authors write as follows. We can transform the
trend filtering problem in
\begin{equation*}
\hat{\beta}=\underset{\beta \in \mathbb{R}^n}{\operatorname{argmin}} \frac{1}{2}\|y-\beta\|_2^2+\frac{n^k}{k!} \lambda\left\|D^{(k+1)} \beta\right\|_1,
\end{equation*}
into lasso form, just like the representation for locally adaptive regression
splines in Lemma 1 (not reproduced for brevity).

\paragraph{Lemma 2.} The trend filtering problem
above is
equivalent to the lasso problem:
$$
\hat{\alpha}=\underset{\alpha \in \mathbb{R}^n}{\operatorname{argmin}} \frac{1}{2}\|y-H \alpha\|_2^2+\lambda \sum_{j=k+2}^n\left|\alpha_j\right|
$$
in that the solutions satisfy $\hat{\beta}=H \hat{\alpha}$. Here, the predictor matrix $H \in \mathbb{R}^{n \times n}$ is given by
$$
H_{i j}= \begin{cases}i^{j-1} / n^{j-1} & \text { for } i=1, \ldots n, j=1, \ldots k+1, \\ 0 & \text { for } i \leq j-1, j \geq k+2 \\ \sigma_{i-j+1}^{(k)} \cdot k!/ n^k & \text { for } i>j-1, j \geq k+2\end{cases}
$$
where we define $\sigma_i^{(0)}=1$ for all $i$ and
$$
\sigma_i^{(k)}=\sum_{j=1}^i \sigma_j^{(k-1)} \text { for } k=1,2,3, \ldots,
$$
i.e., $\sigma_i^{(k)}$ is the $k$ th order cumulative sum of $(1,1, \ldots 1) \in \mathbb{R}^i$.

%% file: main-arxiv.bbl
\begin{thebibliography}{}

\bibitem[Aghaeepour, 2010]{flowmeans}
Aghaeepour, N. (2010).
\newblock Flowmeans: non-parametric flow cytometry data gating.
\newblock {\em R package version}, 1(0).

\bibitem[Arnold et~al., 2014]{arnold2014glmgen}
Arnold, T., Sadhanala, V., and Tibshirani, R.~J. (2014).
\newblock glmgen: Fast generalized lasso solver.

\bibitem[Arnold et~al., 2020]{arnold2020package}
Arnold, T.~B., Tibshirani, R.~J., Arnold, M.~T., and ByteCompile, T. (2020).
\newblock Package ‘genlasso’.
\newblock {\em Statistics}, 39(3):1335--1371.

\bibitem[Bien and Vossler, 2023]{litr}
Bien, J. and Vossler, P. (2023).
\newblock {\em litr: Literate Programming for Writing R Packages}.
\newblock R package version 0.9.1.

\bibitem[Boyd et~al., 2011]{boyd-admm}
Boyd, S., Parikh, N., Chu, E., Peleato, B., and Eckstein, J. (2011).
\newblock Distributed optimization and statistical learning via the alternating
  direction method of multipliers.
\newblock {\em Foundations and Trends® in Machine Learning}, 3(1):1--122.

\bibitem[Cheung et~al., 2021]{Cheung2021}
Cheung, M., Campbell, J.~J., Whitby, L., Thomas, R.~J., Braybrook, J., and
  Petzing, J. (2021).
\newblock Current trends in flow cytometry automated data analysis software.
\newblock {\em Cytometry Part A}, 99(10):1007–1021.

\bibitem[Dempster et~al., 1977]{Dempster1977}
Dempster, A.~P., Laird, N.~M., and Rubin, D.~B. (1977).
\newblock Maximum likelihood from incomplete data via the em algorithm.
\newblock {\em Journal of the Royal Statistical Society Series B: Statistical
  Methodology}, 39(1):1–22.

\bibitem[Dubelaar et~al., 1999]{Dubelaar1999-af}
Dubelaar, G.~B., Gerritzen, P.~L., Beeker, A.~E., Jonker, R.~R., and Tangen, K.
  (1999).
\newblock Design and first results of {CytoBuoy}: a wireless flow cytometer for
  in situ analysis of marine and fresh waters.
\newblock {\em Cytometry}, 37(4):247--254.

\bibitem[Friedman et~al., 2022]{friedman2022package}
Friedman, J., Hastie, T., Tibshirani, R., Narasimhan, B., Tay, K., Simon, N.,
  and Qian, J. (2022).
\newblock Package ‘glmnet’.
\newblock {\em Journal of Statistical Software. 2010a}, 33(1).

\bibitem[Ge and Sealfon, 2012]{flowpeaks}
Ge, Y. and Sealfon, S.~C. (2012).
\newblock flowpeaks: a fast unsupervised clustering for flow cytometry data via
  k-means and density peak finding.
\newblock {\em Bioinformatics}, 28(15):2052--2058.

\bibitem[Huang et~al., 2013]{Huang2013}
Huang, M., Li, R., and Wang, S. (2013).
\newblock Nonparametric mixture of regression models.
\newblock {\em Journal of the American Statistical Association},
  108(503):929–941.

\bibitem[Hyun et~al., 2023]{hyun2020modeling}
Hyun, S., Rolf~Cape, M., Ribalet, F., and Bien, J. (2023).
\newblock Modeling cell populations measured by flow cytometry with covariates
  using sparse mixture of regressions.
\newblock {\em The Annals of Applied Statistics}, 17(1).

\bibitem[Johnson, 2013]{johnson2013dynamic}
Johnson, N.~A. (2013).
\newblock A dynamic programming algorithm for the fused lasso and l
  0-segmentation.
\newblock {\em Journal of Computational and Graphical Statistics},
  22(2):246--260.

\bibitem[Kim et~al., 2009]{kim2009ell_1}
Kim, S.-J., Koh, K., Boyd, S., and Gorinevsky, D. (2009).
\newblock $\backslash$ell\_1 trend filtering.
\newblock {\em SIAM review}, 51(2):339--360.

\bibitem[Knuth, 1992]{Knuth1992-zz}
Knuth, D.~E. (1992).
\newblock {\em Literate Programming}.
\newblock Center for the Study of Language and Information Publication Lecture
  Notes. Centre for the Study of Language \& Information, Stanford, CA.

\bibitem[Kuhn, 1955]{kuhn1955hungarian}
Kuhn, H.~W. (1955).
\newblock The hungarian method for the assignment problem.
\newblock {\em Naval research logistics quarterly}, 2(1-2):83--97.

\bibitem[Malek et~al., 2014]{flowdensity}
Malek, M., Taghiyar, M.~J., Chong, L., Finak, G., Gottardo, R., and Brinkman,
  R.~R. (2014).
\newblock flowdensity: reproducing manual gating of flow cytometry data by
  automated density-based cell population identification.
\newblock {\em Bioinformatics}, 31(4):606–607.

\bibitem[Minoura et~al., 2019]{cybertrack}
Minoura, K., Abe, K., Maeda, Y., Nishikawa, H., and Shimamura, T. (2019).
\newblock Model-based cell clustering and population tracking for time-series
  flow cytometry data.
\newblock {\em BMC Bioinformatics}, 20(S23).

\bibitem[Olson et~al., 2003]{Olson2003}
Olson, R.~J., Shalapyonok, A., and Sosik, H.~M. (2003).
\newblock An automated submersible flow cytometer for analyzing pico- and
  nanophytoplankton: Flowcytobot.
\newblock {\em Deep Sea Research Part I: Oceanographic Research Papers},
  50(2):301–315.

\bibitem[Quintelier et~al., 2021]{flowsom}
Quintelier, K., Couckuyt, A., Emmaneel, A., Aerts, J., Saeys, Y., and
  Van~Gassen, S. (2021).
\newblock Analyzing high-dimensional cytometry data using flowsom.
\newblock {\em Nature Protocols}, 16(8):3775–3801.

\bibitem[Ramdas and Tibshirani, 2016]{ramdas2016fast}
Ramdas, A. and Tibshirani, R.~J. (2016).
\newblock Fast and flexible admm algorithms for trend filtering.
\newblock {\em Journal of Computational and Graphical Statistics},
  25(3):839--858.

\bibitem[Rand, 1971]{rand-index}
Rand, W.~M. (1971).
\newblock Objective criteria for the evaluation of clustering methods.
\newblock {\em Journal of the American Statistical Association},
  66(336):846--850.

\bibitem[Ribalet et~al., 2019]{ribalet2019seaflow}
Ribalet, F., Berthiaume, C., Hynes, A., Swalwell, J., Carlson, M., Clayton, S.,
  Hennon, G., Poirier, C., Shimabukuro, E., White, A., et~al. (2019).
\newblock Seaflow data v1, high-resolution abundance, size and biomass of small
  phytoplankton in the north pacific.
\newblock {\em Scientific Data}, 6(1):1--8.

\bibitem[Sosik et~al., 2010]{Sosik2010}
Sosik, H.~M., Olson, R.~J., and Armbrust, E.~V. (2010).
\newblock {\em Flow Cytometry in Phytoplankton Research}, page 171–185.
\newblock Springer Netherlands.

\bibitem[Steidl et~al., 2006]{steidl2006splines}
Steidl, G., Didas, S., and Neumann, J. (2006).
\newblock Splines in higher order tv regularization.
\newblock {\em International journal of computer vision}, 70(3):241--255.

\bibitem[Swalwell et~al., 2011]{seaflow-paper}
Swalwell, J.~E., Ribalet, F., and Armbrust, E.~V. (2011).
\newblock Seaflow: A novel underway flow‐cytometer for continuous
  observations of phytoplankton in the ocean.
\newblock {\em Limnology and Oceanography: Methods}, 9(10):466–477.

\bibitem[Tibshirani et~al., 2005]{tibshirani2005sparsity}
Tibshirani, R., Saunders, M., Rosset, S., Zhu, J., and Knight, K. (2005).
\newblock Sparsity and smoothness via the fused lasso.
\newblock {\em Journal of the Royal Statistical Society Series B: Statistical
  Methodology}, 67(1):91--108.

\bibitem[Tibshirani, 2014]{tibshirani2014adaptive}
Tibshirani, R.~J. (2014).
\newblock Adaptive piecewise polynomial estimation via trend filtering.
\newblock {\em The Annals of Statistics}, 42(1):285--323.

\bibitem[Wu and Yao, 2016]{mixquantreg}
Wu, Q. and Yao, W. (2016).
\newblock Mixtures of quantile regressions.
\newblock {\em Computational Statistics \& Data Analysis}, 93:162--176.

\bibitem[Xiang and Yao, 2017]{xiangyao2017}
Xiang, S. and Yao, W. (2017).
\newblock Semiparametric mixtures of regressions with single-index for model
  based clustering.
\newblock {\em Advances in Data Analysis and Classification}, 14.

\bibitem[Xiang et~al., 2019]{Xiang2019}
Xiang, S., Yao, W., and Yang, G. (2019).
\newblock An overview of semiparametric extensions of finite mixture models.
\newblock {\em Statistical Science}, 34(3).

\bibitem[Xie, 2016]{bookdown}
Xie, Y. (2016).
\newblock {\em bookdown: Authoring Books and Technical Documents with {R}
  Markdown}.
\newblock Chapman and Hall/CRC, Boca Raton, Florida.

\bibitem[Yoo et~al., 2003]{slurm}
Yoo, A.~B., Jette, M.~A., and Grondona, M. (2003).
\newblock Slurm: Simple linux utility for resource management.
\newblock In Feitelson, D., Rudolph, L., and Schwiegelshohn, U., editors, {\em
  Job Scheduling Strategies for Parallel Processing}, pages 44--60, Berlin,
  Heidelberg. Springer Berlin Heidelberg.

\end{thebibliography}
